\documentclass[a4paper,12pt]{article}
\usepackage{amsmath,amssymb}
\usepackage{axodraw}
\begin{document}
\thispagestyle{empty}

\begin{flushright}
JINR E2-80-483\\
\end{flushright}

\vspace{3.cm}

\begin{center}
\large\textbf{THREE-LOOP CALCULATIONS IN \\[.1cm]
NON-ABELIAN GAUGE THEORIES}
\end{center}

\vspace{.2cm}

\begin{center}
\textsc{O.V.\,Tarasov} \,and \,\textsc{A.A.Vladimirov} \\[.1cm]
\textit{JINR, Dubna} \\[.2cm]
\texttt{e-mail: otarasov@jinr.ru, alvladim@theor.jinr.ru}

\end{center}

\vspace{2cm}

\begin{center}
\large\textbf{Abstract}
\end{center}

A detailed description of the method for analytical
evaluation of the three-loop contributions to renormalization
group functions is presented. This method is
employed to calculate the charge renormalization
function and anomalous dimensions for non-Abelian gauge theories
with fermions in the three-loop approximation.
A three-loop expression for the effective charge of QCD is given.
Charge renormalization effects in the $SU(4)$-supersymmetric gauge
model is shown to vanish at this level.
A complete list of required formulas is given in Appendix.

The above-mentioned results of three-loop calculations have been published
by the present authors (with A.Yu.\,Zharkov and L.V.\,Avdeev) in 1980
in Physics Letters B. The present text, which treats the subject in more
details and contains a lot of calculational techniques, has also been
published in 1980 as the JINR Communication E2-80-483.

\newpage

\begin{center}
\large\textbf{1. \,Introduction}
\end{center}

The renormalization group method when applied to asymptotically
free models results in an ``improved'' perturbation theory.
Its expansion parameter, an effective charge
$\bar{g}^2(Q^2\!/\Lambda^2,g^2)$, decreases logarithmically with the increase
in the momentum transfer $Q^2$. The existent QCD calculations of various
deep inelastic processes in the first two orders in $\bar{g}^2$
appear to be consistent with the present experimental data [1].
However, the next-to-leading corrections (i.e., those $\sim \bar{g}^4$ )
are fairly large. It leaves open the possibility that the higher-order
contributions will be important.

 The calculations in higher orders are also of interest
from another standpoint.  They might serve us a starting
point for summing the perturbation theory expansions of QCD, as
it is done, for instance, in the $\phi^4$ model [2].
Moreover, these calculations can shed light on some peculiar aspects of certain
field theory models. For example, in the $SU(4)$-supersymmetric
non-Abelian gauge model derived in [3,4],
the charge renormalization effects are shown to vanish to the two-loop
order [5].
The corresponding three-loop calculations presented
below give the same answer: The charge renormalization function
$\beta(g^2)$ is equal to zero. Apparently, the vanishing of
$\beta(g^2)$ at the three-loop level is not a sheer coincidence, but
an indication that this effect holds to all orders.

The first three-loop QCD calculation in the framework of the
renormalization group has been performed in [6],
where the total cross section of the $e^+e^-$ - annihilation into hadrons
has been computed analytically. This result is confirmed in [7] by
a numerical calculation and in [8] also analytically.
However, these calculations involve the $\beta(g^2)$ function to order $g^6$,
whereas all other three-loop QCD calculations require the next,
$\sim\!g^8$, contribution to $\beta(g^2)$. The charge
renormalization function $\beta(g^2)$ for the non-Abelian gauge theory
including fermions is known to $g^6$ only, i.e., in the two-loop
approximation [9]. In the present paper we describe a method
which enables one to evaluate $\beta(g^2)$ at the three-loop level.
We present the results of these calculations and the full list of
needed formulas.

\vspace{.3cm}

\begin{center}
\large\textbf{2. \,Renormalization group in the minimal subtraction scheme}
\end{center}

We consider a non-Abelian gauge group theory with
fermions belonging to the representation R of the gauge
group G:
\begin{equation}
\label{Lagrangian}
{\cal L} = - \frac{1}{4} G^{a}_{\mu \nu} G^{a}_{\mu \nu}
-\frac{1}{2\alpha} \left( \partial_{\mu} A^a_{\mu} \right)^2
-\partial_{\mu} \bar{\eta}^a \partial_{\mu} {\eta}^a
+g f^{abc}\bar{\eta}^a A_{\mu}^b \partial_{\mu}
\eta^c + i \sum_{m=1}^f \bar{\psi}_i^{~m}
\hat{\cal{D} }\psi_i^m,
\end{equation}
$$ G^{a}_{\mu \nu}=\partial_{\mu} A^a_{\nu}-\partial_{\nu} A^a_{\mu}
+g f^{abc} A_{\mu}^b A_{\nu}^c\,, \ \ \ \ \ \ \ \ \
{\cal{D}}_{\mu}\psi^m_i = \partial_{\mu}\psi_i^m
-igR_{ij}^a\psi^m_j A_{\mu}^a\,.
$$
Here $\eta^a$ is the ghost field, $\alpha$ is the gauge
parameter, and $ f^{abc}$ are the totally antisymmetric structure
constants of the gauge group $G$. The indices of the fermion field
 $\psi^m_i$ specify color $(i)$ and flavor $(m)$, respectively.
The matrices $R^a$ obey the following relations:
\begin{equation}
\label{equation2}
[R^a,R^b]_- = if^{abc}R^c, \ \ \ \ f^{acd}f^{bcd} = \,C_A\delta^{ab},
\ \ \ \ R^aR^a = \,C_F I, \ \ \ \ {\rm tr}(R^aR^b) = \,T \delta^{ab}.
\end{equation}
In particular, the values of group invariants $C_A$, $C_F$ and
$T$ in the fundamental (quark) representation of $SU(N)$ are:
\begin{equation}
C_A=N,~~~C_F=\frac{N^2-1}{2N},~~~T=\frac12\ .
\end{equation}

The underlying gauge symmetry of the Lagrangian (\ref{Lagrangian})
  gives rise to the well-known Slavnov-Taylor identities
[10] extensively used throughout the paper. In particular,
a transversality of the radiative corrections to the gluon
propagator allows one to compute such a correction in the scalar
form, i.e., with its Lorentz indices contracted.

We now turn to a brief discussion of the renormalization procedure.
In this paper we adopt the renormalization prescription by 't Hooft
[11], the so-called ``minimal subtraction scheme",
which by definition
subtracts only pole parts in $\varepsilon$ from a given diagram.
 The renormalization constants $Z_{\Gamma}$ relating the
dimensionally regularized 1PI Green function with the renormalized
one,
\begin{equation}
\Gamma_R\left(\frac{Q^2}{\mu^2},\alpha,g^2\right)=\lim_{\varepsilon
\rightarrow 0} Z_{\Gamma}\left(\frac{1}{\varepsilon},\alpha,g^2\right)
\Gamma\left(Q^2,\alpha_B,g_B^2,\varepsilon \right),
\label{gammaR}
\end{equation}
look in this scheme like
\begin{equation}
Z_{\Gamma}\left(\frac{1}{\varepsilon},\alpha,g^2\right)
=1+ \sum_{n=1}^{\infty}c_{\Gamma}^{(n)}(\alpha,g^2)
\varepsilon^{-n},
\label{Zgamma}
\end{equation}
with $\varepsilon = \frac{4-d}{2}$, $d$ being the space-time
dimension. In (\ref{gammaR}) $\mu$ is the renormalization parameter
with the dimension of mass.
The bare charge $g_B^2$ is to be constructed from appropriate
$Z$'s. The most convenient choice is as follows:
\begin{equation}
g_B^2=\mu^{2\varepsilon}g^2 \tilde{Z}^2_1 Z_3^{-1}
\tilde{Z}_3^{-2}.
\label{bare_charge}
\end{equation}
Here $\tilde{Z}_1$ is the renormalization constant of the
ghost-ghost-gluon vertex, $Z_3$ and $\tilde{Z}_3$ being
those of inverted gluon and ghost propagators, respectively.
Note also $\alpha_B$ in (\ref{gammaR}) to be given by $\alpha_B=\alpha Z_3$.
The Green function  $\Gamma_R\left(\frac{Q^2}{\mu^2},\alpha,g^2\right)$
satisfies the renormalization group equation
\begin{equation}
\left[Q^2 \frac{\partial}{\partial Q^2}
-\beta(g^2)\frac{\partial}{\partial g^2}
-\gamma_3(\alpha,g^2)\alpha \frac{\partial}{\partial \alpha}
-\gamma_{\Gamma}(\alpha,g^2) \right]
\Gamma_R\left(\frac{Q^2}{\mu^2},\alpha,g^2\right) =0
\label{RGequation}
\end{equation}
and the normalization condition
 $\Gamma_R\left(\frac{Q^2}{\mu^2},\alpha,0\right)=1$.
The anomalous dimensions $\gamma_{\Gamma}$ are given by the
relation
\begin{equation}
\gamma_{\Gamma}(\alpha,g^2)=g^2 \frac{\partial}{\partial g^2}
c_{\Gamma}^{(1)}(\alpha,g^2).
\label{GammaGamma}
\end{equation}
Similarly, from
\begin{equation}
g^2_B=\mu^{2\varepsilon}\left[ g^2+ \sum_{n=1}^{\infty}
a^{(n)}(g^2)\varepsilon^{-n} \right]
\label{Bare_Charge_ren}
\end{equation}
one obtains the charge renormalization function $\beta$,
\begin{equation}
\beta(g^2) \equiv \left(g^2 \frac{\partial}{\partial g^2}
-1\right ) a^{(1)}(g^2)=
g^2\left[2\tilde{\gamma}_1(\alpha,g^2)
- \gamma_3(\alpha,g^2)-2\tilde{\gamma}_3(\alpha , g^2)  \right],
\label{beta_function}
\end{equation}
which is known to be gauge independent [12].
Thus, the computation of $\gamma_{\Gamma}(\alpha,g^2)$ and
$\beta(g^2)$ requires the functions $c_{\Gamma}^{(1)}(\alpha,g^2)$
for the renormalization constants in the right-hand side
of (\ref{bare_charge}).

The residues of higher-order poles in the expansion
(\ref{Zgamma}) and (\ref{Bare_Charge_ren}) are related with
$c^{(1)}$ and $a^{(1)}$ by the equalities
\begin{equation}
\left[ \beta(g^2)\frac{\partial}{\partial g^2}
+\gamma_3(\alpha, g^2) \alpha \frac{\partial }{\partial \alpha}
+ \gamma_{\Gamma}(\alpha, g^2)\right]
c^{(n)}_{\Gamma}(\alpha,g^2)
=g^2 \frac{\partial}{\partial g^2} c_{\Gamma}^{(n+1)}
(\alpha,g^2),
\label{equation11}
\end{equation}
\begin{equation}
\beta(g^2)\frac{\partial}{\partial g^2}a^{(n)}(g^2)
=\left(g^2 \frac{\partial}{\partial g^2}-1\right)a^{(n+1)}(g^2).
\end{equation}
We choose to work in the Feynman gauge $\alpha=1$ throughout this
paper. For checking the higher residues by means of (\ref{equation11})
one may use the results of the corresponding two-loop calculations
[13] performed in a general gauge.

According to the minimal subtraction prescription [11],
the renormalization constants are uniquely determined by requiring
that all the divergences in $\varepsilon$ disappear from the
product $Z_{\Gamma}\left(\frac{1}{\varepsilon},\alpha,g^2\right)
\Gamma\left(Q^2,\alpha_B,g_B^2,\varepsilon \right)$, so that
the limit $\varepsilon \rightarrow 0$ in (\ref{gammaR}) does
exist. However, we find a somewhat different (but equivalent)
definition [14] to be more convenient:
\begin{equation}
Z_{\Gamma}=1- {{\cal{K}}}R'\Gamma .
\label{equation13}
\end{equation}
An operator ${{\cal{K}}}$ picks out all the pole terms in $\varepsilon$,
\begin{equation}
{{\cal{K}}} \sum_{n}b_n\varepsilon^n=\sum_{n<0}b_n\varepsilon^n.
\label{equation14}
\end{equation}
$R'$ is the BPHZ minimal subtraction procedure
($R$-operation) with its final subtraction missing:
 $R=(1- {{\cal{K}}})R'$. In other words,
the $R'$-operation subtracts all the divergences of internal
subgraphs but does not subtract an overall divergence of a diagram.
To construct $R'$ explicitly one can employ the following recursion
relation [15]:
\begin{equation}
R'G = G + \sum \left(-{{\cal{K}}} R'G_1 \right)\cdot...\cdot
\left(-{{\cal{K}}} R'G_m \right)\cdot G/(G_1+...+G_m)\,,
\label{equation15}
\end{equation}
where the sum is over all sets of disjoint $1PI$ divergent subgraphs of
the diagram $G$, and $G/(G_1+...+G_m)$ is the diagram
obtained from $G$ by contracting $G_1, ...,G_m$ to points
(as an example see Fig.1).



\begin{center}
\begin{picture}(450,140)(0,0)
\Text(322,120)[]{${  KR'}$ }
\Text(7,120)[]{${  R'}$ }
\Text(297,120)[]{${ -2}$ }
\Text(161,120)[]{${ - }$ }

\Text(81,120)[]{${ = }$ }

\Line(338,120)(368,140)
\Line(338,120)(368,105)

\Line(306,100)(306,145)
\Line(306,100)(313,100)
\Line(306,145)(313,145)

\Photon(363,136)(363,107){2}{3}

\Photon(331,120)(338,120){2}{2}

\Photon(347,125)(361,120){2}{2}

\Line(375,100)(375,145)
\Line(375,100)(370,100)
\Line(375,145)(370,145)

\CArc(407,120)(20,0,180)
\CArc(407,120)(20,180,360)


\Photon(377,120)(387,120){2}{2}
\Photon(427,120)(437,120){2}{2}

\Text(182,120)[]{${  KR'}$ }

\Photon(197,120)(213,135){2}{3}
\Photon(213,110)(197,120){2}{3}

\Line(166,100)(166,145)
\Line(166,100)(173,100)
\Line(166,145)(173,145)

\Line(213,140)(213,105)

\Photon(191,120)(196,120){2}{1}

\Line(222,100)(222,145)
\Line(222,100)(217,100)
\Line(222,145)(217,145)


\CArc(255,120)(20,0,180)
\CArc(255,120)(20,180,360)
\Photon(255,140)(255,100){2}{5}
\Photon(225,120)(235,120){2}{2}
\Photon(275,120)(285,120){2}{2}


\CArc(42,120)(20,0,180)
\CArc(42,120)(20,180,360)

\Photon(12,120)(22,120){2}{2}
\Photon(62,120)(72,120){2}{2}

\Photon(42,140)(42,120){2}{4}

\Photon(42,120)(26,108){2}{4}

\Photon(42,120)(60,108){2}{4}

\Vertex(26,108){1.5}
\Vertex(58,108){1.5}

\Vertex(42,120){1.5}
\Vertex(42,140){1.5}

\CArc(117,120)(20,0,180)
\CArc(117,120)(20,180,360)

\Photon(87,120)(97,120){2}{2}
\Photon(137,120)(147,120){2}{2}

\Photon(117,140)(117,120){2}{4}

\Photon(117,120)(101,108){2}{4}

\Photon(117,120)(135,108){2}{4}

\Vertex(101,108){1.5}
\Vertex(133,108){1.5}

\Vertex(117,120){1.5}
\Vertex(117,140){1.5}

%
\Text(9,50)[]{${  R'}$ }

\Line(23,50)(53,70)
\Line(23,50)(53,35)

\Photon(48,66)(48,37){2}{3}
\Photon(16,50)(23,50){2}{2}
\Photon(33,55)(47,50){2}{2}

\Text(62,50)[]{${=}$ }
\Text(122,50)[]{${-}$ }
\Text(232,50)[]{${,}$ }

\Line(78,50)(108,70)
\Line(78,50)(108,35)

\Photon(103,66)(103,37){2}{3}
\Photon(71,50)(78,50){2}{2}
\Photon(87,55)(101,50){2}{2}

\Line(208,50)(228,70)
\Line(208,50)(228,35)

\Photon(223,65)(223,40){2}{3}
\Photon(201,50)(208,50){2}{2}

\Line(133,30)(133,75)
\Line(133,30)(140,30)
\Line(133,75)(140,75)

\Text(149,50)[]{${  KR'}$ }

\Photon(168,50)(183,65){2}{3}
\Photon(183,35)(168,50){2}{3}
\Photon(161,50)(168,50){2}{2}

\Line(183,70)(183,30)

\Line(192,30)(192,75)
\Line(192,30)(187,30)
\Line(192,75)(187,75)

\Text(305,50)[]{${  R'}$ }
\Text(350,50)[]{${ = }$ }

\Photon(318,50)(333,65){2}{3}
\Photon(333,35)(318,50){2}{3}
\Photon(311,50)(318,50){2}{2}

\Line(333,70)(333,30)

\Photon(368,50)(383,65){2}{3}
\Photon(383,35)(368,50){2}{3}
\Photon(361,50)(368,50){2}{2}

\Line(383,70)(383,30)
\Text(140,0)[]{${\rm~~~~~~~~~~~~~~~~~~~~~~~Fig.~1}$ }
\label{Fig1}
\end{picture}

\end{center}

The ${{\cal{K}}} R'G$ is the negative of a contribution from
$G$ to an appropriate renormalization constant. The computation
of ${{\cal{K}}} R'G$ is simplified drastically owing to the following
fact [16].

Let a diagram $G$ be infrared finite in a range of external
momenta $k_i$ and internal masses $m_j$. Then in this range
${{\cal{K}}} R'G$ is a polynomial in $k_i$ and $m_j$. Therefore,
it either is independent of $k_i$ and $m_j$ (for a logarithmically
divergent diagram $G$) or loses such a dependence after differentiating
once or twice with respect to $k_i$.

\vspace{.3cm}

\begin{center}
\large\textbf{3. \,A method for computing three-loop integrals}
\end{center}

This feature of ${{\cal{K}}} R'G$ provides the basis for a simple
and efficient computational technique developed in [15],
which enables one to evaluate analytically all three-loop contributions
to the renormalization group functions $\gamma$ and $\beta$ in any
renormalizable theory. It is shown in [15] that one may
calculate ${{\cal{K}}} R'G$ (properly differentiated, if necessary) with
all its external momenta equal to zero and with an auxiliary mass
$m\neq 0$ introduced into one of its internal lines (which is
sufficient to remove all infrared divergences). The momentum integration
corresponding to this line is chosen to be the last one. It looks like
\begin{equation}
\int \frac{dp}{(p^2)^{\alpha}(p^2+m^2)}
\label{equation16}
\end{equation}
and is readily done using Eq. (\ref{equationA8}) in Appendix.
We thus show the last momentum integration to be trivial. Therefore, the
problem of three-loop calculations reduces to computing the
two-loop massless integrals depending on a single momentum
$p^2$,
\begin{equation}
\int \frac{dt~dq}{t^{2\alpha} q^{2\beta} (p-t)^{2\gamma}
(p-q)^{2\sigma} (t-q)^{2\rho}}
\label{equation17}
\end{equation}
with $\alpha$, $\beta$, $\gamma$, $\sigma$ and $\rho$ being integers. If one
of the denominators is missing (e.g., $\rho = 0, -1, -2, ...$ ) the integral
(\ref{equation17}) can be evaluated by sequential use of
Eq. (\ref{equationA9}). Otherwise one needs the non-trivial two-loop
integration formulas deduced in [17]
through the $x$-space Gegenbauer polynomial technique.
In Appendix we give a list of relevant integrals of the type
(\ref{equation17}).

As an illustrative example we consider an integral
{
\begin{equation}
{{ J}=\int \frac{dp~ dq~ dt~(qt)^2}
{p^2q^2t^2(p-q)^2(p-t)^2(k-q)^2(k-t)^2}}
\label{equation18}
\end{equation}
}
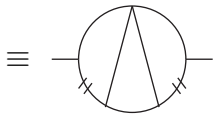
\begin{picture}(50,40)(0,30)
\Text(160,50)[]{${ \equiv}$ }
\CArc(200,50)(20,0,180)
\CArc(200,50)(20,180,360)
\Line(170,50)(180,50)
\Line(220,50)(230,50)

\Line(200,70)(190,32)
\Line(200,70)(210,32)

\Line(182,37)(185,41)
\Line(180,39)(183,43)

\Line(215,41)(218,37)
\Line(217,43)(220,39)

\end{picture}
\vspace{1.0cm}

Due to quadratic divergence, it should be differentiated
twice with respect to $k$. Using the relation
\begin{equation}
\frac{\partial^2}{\partial k_{\mu} ~\partial k_{\mu}}
\left[\frac{1}{(k-q)^2 (k-t)^2} \right]=
\frac{8(k-q)(k-t)+4\varepsilon [(k-q)^2 +(k-t)^2] }
{(k-q)^4(k-t)^4}
\label{equation19}
\end{equation}
we obtain ${{\cal{K}}} \partial^2 R'J $ as displayed in Fig.2 in
self-evident notation. Since
${{\cal{K}}}  R'J = k^2A\left(\frac{1}{\varepsilon}\right)$,
we finally get
\begin{equation}
A\left(\frac{1}{\varepsilon} \right) =
 {{\cal{K}}} \frac{1}{8-4\varepsilon}
{{\cal{K}}} \partial^2 R'J =
(i\pi^2)^3
\left( \frac{1}{24 \varepsilon^2}+\frac{1}{32\varepsilon}\right).
\label{equation20}
\end{equation}
The last two diagrams in Fig.2 diverge logarithmically so that
one can compute them with $k=0$ provided that a non-zero
mass is introduced into one of the differentiated lines, i.e.,
into that with a blob.
%
%
\begin{center}
\begin{picture}(450,150)(0,0)

\Text(68,122)[]{${ \partial^2}$ }
\CArc(100,120)(15,0,180)
\CArc(100,120)(15,180,360)
\Line(75,120)(85,120)
\Line(115,120)(125,120)
\Line(100,105)(92,132)
\Line(100,105)(108,132)

\Line(89,125)(84,129)
\Line(88,123)(83,126)

\Line(111,125)(116,129)
\Line(112,123)(117,126)


\Text(159,122)[]{${=~ 8}$ }
\CArc(200,120)(15,0,180)
\CArc(200,120)(15,180,360)
\Line(175,120)(185,120)
\Line(215,120)(225,120)
\Line(200,105)(192,132)
\Line(200,105)(208,132)

\Line(189,125)(184,129)
\Line(188,123)(183,126)

\Line(211,125)(216,129)
\Line(212,123)(217,126)

\Vertex(188,111){1.5}
\Vertex(212,111){1.5}
\Line(193,110)(189,106)

\Line(207,110)(211,106)


\Text(256,122)[]{${+~ 8 ~\varepsilon}$ }
\Text(335,119)[]{${,}$ }
\CArc(300,120)(15,0,180)
\CArc(300,120)(15,180,360)
\Line(275,120)(285,120)
\Line(315,120)(325,120)
\Line(300,105)(292,132)
\Line(300,105)(308,132)

\Line(289,125)(284,129)
\Line(288,123)(283,126)

\Line(311,125)(316,129)
\Line(312,123)(317,126)

\Vertex(288,111){1.5}


\Text(48,72)[]{${R'}$ }
\CArc(80,70)(15,0,180)
\CArc(80,70)(15,180,360)
\Line(55,70)(65,70)
\Line(95,70)(105,70)
\Line(80,55)(72,82)
\Line(80,55)(88,82)

\Line(69,75)(64,79)
\Line(68,73)(63,76)

\Line(91,75)(96,79)
\Line(92,73)(97,76)

\Text(125,70)[]{${=}$ }
\CArc(160,70)(15,0,180)
\CArc(160,70)(15,180,360)
\Line(135,70)(145,70)
\Line(175,70)(185,70)
\Line(160,55)(152,82)
\Line(160,55)(168,82)

\Line(149,75)(144,79)
\Line(148,73)(143,76)

\Line(171,75)(176,79)
\Line(172,73)(177,76)
\Line(210,55)(210,85)
\Line(210,85)(213,85)
\Line(210,55)(213,55)

\Text(200,70)[]{${-2}$ }
\Text(220,70)[]{${\cal{K}}$ }
\Line(220,80)(250,80)
\Line(225,80)(239,58)
\Line(245,80)(231,58)
\Line(233,83)(233,77)
\Line(237,83)(237,77)
\Line(255,55)(255,85)
\Line(255,85)(252,85)
\Line(255,55)(252,55)

\CArc(278,70)(15,0,180)
\CArc(278,70)(15,180,360)
\CArc(263,85)(15,-90,0)

\Line(258,70)(263,70)
\Line(293,70)(298,70)
\Line(292,83)(286,78)
\Line(293,81)(288,76)
\Text(313,70)[]{${-2}$ }
\Line(321,55)(321,85)
\Line(321,85)(324,85)
\Line(321,55)(324,55)

\Text(339,70)[]{${\cal{K}}R' $ }
\Line(340,80)(380,80)
\Line(360,80)(360,55)
\Line(345,80)(360,60)
\Line(375,80)(360,60)

\Line(355,83)(355,77)
\Line(353,83)(353,77)

\Line(385,85)(385,55)
\Line(382,85)(385,85)
\Line(382,55)(385,55)

\CArc(410,70)(15,0,180)
\CArc(410,70)(15,180,360)

\Line(390,70)(395,70)
\Line(425,70)(430,70)

\Line(408,88)(408,82)
\Line(412,88)(412,82)
\Text(437,69)[]{${,}$ }

\Text(58,22)[]{${{\cal{K}} \partial^2 R'}$ }
\CArc(100,20)(15,0,180)
\CArc(100,20)(15,180,360)
\Line(75,20)(85,20)
\Line(115,20)(125,20)
\Line(100,5)(92,32)
\Line(100,5)(108,32)

\Line(89,25)(84,29)
\Line(88,23)(83,26)

\Line(111,25)(116,29)
\Line(112,23)(117,26)


\Line(156,35)(159,35)
\Line(156,5)(156,35)
\Line(156,5)(159,5)

\Text(144,22)[]{${=~ {\cal{K} }}$ }

\Text(175,22)[]{${ 8{\cal{K} } R'}$ }

\CArc(210,20)(15,0,180)
\CArc(210,20)(15,180,360)
\Line(185,20)(195,20)
\Line(225,20)(235,20)
\Line(210,5)(202,32)
\Line(210,5)(218,32)

\Line(199,25)(194,29)
\Line(198,23)(193,26)

\Line(221,25)(226,29)
\Line(222,23)(227,26)

\Vertex(198,11){1.5}
\Vertex(222,11){1.5}
\Line(203,10)(199,6)

\Line(217,10)(221,6)


\Text(265,22)[]{${ + 8~ \varepsilon~ {\cal{K}} R' }$ }

\Line(348,35)(345,35)
\Line(348,5)(348,35)
\Line(348,5)(345,5)
\Text(355,19)[]{${.}$ }

\CArc(315,20)(15,0,180)
\CArc(315,20)(15,180,360)
\Line(290,20)(300,20)
\Line(330,20)(340,20)
\Line(315,5)(307,32)
\Line(315,5)(323,32)

\Line(304,25)(299,29)
\Line(303,23)(298,26)

\Line(326,25)(331,29)
\Line(327,23)(332,26)

\Vertex(303,11){1.5}
\end{picture}
\end{center}
\begin{center}
Fig.2
\end{center}
 The problem of evaluating ${{\cal{K}}} R'G$
at the three-loop level thus reduces to the integrations
(\ref{equation16}) and (\ref{equation17}). The described procedure
has been employed in a considerable part of the calculations
presented in this paper.

One can also determine the pole part of (\ref{equation18}), ${{\cal{K}}}J$,
by means of a somewhat different method, which involves transferring
an external momentum to the other vertex in order to simplify the
denominator.
\begin{center}
\begin{picture}(200,40)(0,0)
\CArc(80,20)(15,0,180)
\CArc(80,20)(15,180,360)
\Line(55,20)(65,20)
\Line(95,20)(105,20)
\Line(80,5)(72,32)
\Line(80,5)(88,32)

\Line(69,25)(64,29)
\Line(68,23)(63,26)

\Line(91,25)(96,29)
\Line(92,23)(97,26)

\Text(125,20)[]{${-}$ }
\CArc(160,20)(15,0,180)
\CArc(160,20)(15,180,360)

\Vertex(145,20){1.5}
\Line(175,20)(185,20)

\Line(160,5)(152,32)
\Line(160,5)(168,32)
\Line(160,5)(160,0)

\Line(149,25)(144,29)
\Line(148,23)(143,26)

\Line(171,25)(176,29)
\Line(172,23)(177,26)
\end{picture}
\end{center}
\begin{center}
Fig.3
\end{center}
 Consider the difference (Fig.3)
\begin{eqnarray}
&&\int \frac{dp~ dq ~dt ~(qt)^2}{p^2q^2t^2(p-q)^2(p-t)^2(k-q)^2(k-t)^2}
\left[1- \frac{(k-q)^2}{q^2} \right]
\nonumber
\\
&&~~~=\int \frac{dp~ dq ~dt ~(qt)^2(2k_{\mu}q_{\mu}-k^2 )}
{p^2t^2q^4(p-q)^2(p-t)^2(k-q)^2(k-t)^2}
\equiv 2 k_{\mu} J_{\mu}-k^2 J_1\,.
\label{equation21}
\end{eqnarray}
Let us further subtract from $J_{\mu}$ the other integral
having a more simple structure of the denominator:
\begin{equation}
J_{\mu}-\!\int\!\frac{dp ~dq ~dt ~q_{\mu} (qt)^2}
{p^2t^4q^4(p-q)^2(p-t)^2(k-q)^2 } =
\!\int\!\frac{dp~ dq ~dt ~q_{\mu} (qt)^2[2k_{\nu}t_{\nu}-k^2]}
{p^2q^4t^4(p-q)^2(p-t)^2(k-q)^2(k-t)^2}.
\label{equation22}
\end{equation}
There is only one (logarithmically) divergent integral in the
right-hand side of (\ref{equation22}), namely
\begin{equation}
\int \frac{dp~dq ~dt ~2q_{\mu} t_{\nu}(qt)^2 }
{p^2 q^4 t^4 (p-q)^2(p-t)^2 (k-q)^2(k-t)^2}.
\label{equattion23}
\end{equation}
Due to the absence of divergent subgraphs, its pole part does
not depend on $k$ and coincides with
\begin{equation}
{{\cal{K}}} \int \frac{dp~dq~dt~~2q_{\mu} t_{\nu}~(qt)^2}
{p^2q^4 t^6 (p-q)^2 (p-t)^2 (k-q)^2}.
\label{equation24}
\end{equation}
As to the integral $J_1$, it diverges logarithmically and contains
divergent subgraphs. We note the difference
\begin{equation}
J_1 -\int \frac{dp~dq~dt~~(qt)^2}
{p^2q^4 t^4 (p-q)^2 (p-t)^2 (k-q)^2}
\label{equation25}
\end{equation}
to be convergent, and combining the last five relations
finally obtain
\begin{equation}
\label{equation26}
\mathcal{K} J = \mathcal{K}\!\int
\frac{dp~dq~dt~(qt)^2[4(kt)(kq)+2q^2t^2 -t^2(k-q)^2 ]}
{p^2 q^4 t^6 (k-q)^2 (p-q)^2(p-t)^2}
 = -(i\pi^2)^3\!\left(\frac{1}{12\varepsilon^2}
 + \frac{25}{32\varepsilon}\right).
\end{equation}
This integral is easy to evaluate with the use of formulas
listed in Appendix. Adding to (\ref{equation26}) the appropriate counterterms
gives for ${\cal{K}} R'J $ the same answer as in (\ref{equation20}).

The essence of the procedure presented above is as follows.
One subtracts from the initial integral $J$ an infrared
finite integral $J'$ with a more simple denominator reducing
thus the degree of divergence. Such a subtraction is to be
repeated until the difference  becomes convergent.

\vspace{.3cm}

\begin{center}
\large\textbf{4. \,Calculation of specific diagrams}
\end{center}

It is now seen that the three-loop momentum integrals
contributing to $Z'$s are always calculable. However,
one must introduce an auxiliary mass into the diagram (which as a
rule represents a sum of distinct integrals similar to (\ref{equation18}))
and into all its counterterms in a consistent fashion.
For the most complicated diagrams of the gluon propagator this
task appears to be unmanageable. Therefore, we deal with the
diagrams of the topological type, depicted in Fig.4,
as follows.
We reduce the numerator of the integrand to the scalar form
and then decompose it into a sum of invariants like
$k^2(q-t)^4$, $p^2q^2(p-t)^2, ...$. Canceling numerator against
denominator and taking symmetry into account results in at most
66  distinct three-loop massless integrals. Their pole parts are
to be found either by direct use of (\ref{equationA9})-(\ref{equationA14})
or by differentiating, introducing a mass, and then converting
${{\cal{K}}}R'$ into ${{\cal{K}}}$ through the compensating
subtraction. The latter  pole parts are given in Appendix.
\begin{center}
\begin{picture}(250,60)(0,0)
\CArc(30,30)(15,0,180)
\CArc(30,30)(15,180,360)
\Line(10,30)(15,30)
\Line(45,30)(50,30)

\Line(24,43)(24,17)
\Line(36,43)(36,17)

\CArc(90,30)(15,0,180)
\CArc(90,30)(15,180,360)
\Line(70,30)(75,30)
\Line(105,30)(110,30)
\Line(90,45)(90,30)
\Line(90,30)(79,20)
\Line(90,30)(101,20)

\CArc(150,30)(15,0,180)
\CArc(150,30)(15,180,360)
\Line(130,30)(135,30)
\Line(165,30)(170,30)
\Line(150,15)(141,42)
\Line(150,15)(159,42)

\CArc(210,30)(15,0,180)
\CArc(210,30)(15,180,360)
\Line(190,30)(195,30)
\Line(225,30)(230,30)
\Line(210,45)(210,15)
\Line(195,30)(210,30)

\end{picture}
\end{center}
\begin{center}
Fig.4
\end{center}
\begin{center}
\begin{picture}(300,60)(0,0)
\CArc(30,30)(15,0,180)
\CArc(30,30)(15,180,360)
\Line(10,30)(15,30)
\Line(45,30)(50,30)
\CArc(46,30)(22,135,221)

\CArc(14,30)(22,-45,45)

\CArc(90,30)(15,0,180)
\CArc(90,30)(15,180,360)
\Line(70,30)(75,30)
\Line(105,30)(110,30)

\CArc(90,30)(7,0,180)
\CArc(90,30)(7,180,360)
\Line(90,45)(90,37)
\Line(90,23)(90,15)

\CArc(150,30)(15,0,180)
\CArc(150,30)(15,180,360)
\Line(130,30)(135,30)
\Line(165,30)(170,30)
\Line(150,45)(150,15)
\CArc(135,16)(11,2,88)

\CArc(210,30)(15,0,180)
\CArc(210,30)(15,180,360)
\Line(190,30)(195,30)
\Line(225,30)(230,30)
\Line(210,45)(210,15)
\CArc(195,45)(15,-90,0)
\CArc(270,30)(15,0,180)
\CArc(270,30)(15,180,360)
\Line(250,30)(255,30)
\Line(285,30)(290,30)
\CArc(270,53)(15,230,310)
\CArc(270,7)(15,50,130)
\end{picture}
\end{center}
\begin{center}
Fig.5
\end{center}

The propagator diagrams of more simple (``nested") topology
(Fig.5) can be computed straightforwardly using (\ref{equationA9})-(\ref{equationA14}).
The remaining topological type is represented by a single diagram (all
others equal zero owing to the antisymmetry of the group
structure constants) which can be easily calculated by means
of differentiation:
\begin{center}
\begin{picture}(300,45)(0,0)
\Text(10,25)[]{$g_{\mu \nu}$ }
\Text(84,25)[]{$=$ }
\CArc(45,25)(15,0,180)
\CArc(45,25)(15,180,360)
\Photon(20,25)(30,25){2}{2}
\Photon(60,25)(70,25){2}{2}
\Photon(34,35)(56,15){2}{4}
\Photon(34,15)(56,35){2}{4}
\end{picture}
\end{center}
\begin{equation}
=\frac{ g^6 T(C_F-C_A)\left( C_F-\frac{C_A}{2} \right)}
{(4\pi)^6 (k^2)^{3\varepsilon-1}}
\left( \frac{16}{3\varepsilon^2}+\frac{20}{\varepsilon}
-\frac{32}{\varepsilon} \zeta(3)  +O(1) \right).
\label{equation27}
\end{equation}

All the diagrams of the ghost-ghost-gluon vertex diverge
logarithmically. We evaluate them setting all external momenta
to be zero and introducing an auxiliary mass into one of the
internal lines. For each particular diagram this ``potentially
infrared" line is easy to identify.

Thus, all the diagrams of a certain Green function are
calculated in the same fashion: with an auxiliary mass for the
vertices and without it for propagators. It enables one
to perform the subtractions either following 't Hooft
[11] or determining ${{\cal{K}}}R'G$ for
each individual diagram. In order to check the intermediate
results we choose the latter way.

The problem of evaluating the group weights appear to be
of no substantial difficulty. Mostly it reduces to making
contractions in the products of several structure constants
$f^{abc}$. The following graphical representation is here
of great use [18].
%
%
\begin{center}
\begin{picture}(200,35)(0,0)
\Line(30,20)(30,5)
\Line(19,33)(30,20)
\Line(41,33)(30,20)
\Text(67,20)[]{$= f^{abc}~,$ }
\Line(115,20)(150,20)
\Text(175,20)[]{$= \delta^{ab}~,$ }
\end{picture}
\end{center}
\begin{center}
\begin{picture}(250,30)(0,0)
\CArc(30,20)(15,0,180)
\CArc(30,20)(15,180,360)
\Line(8,20)(15,20)
\Line(45,20)(52,20)
\Text(80,20)[]{$=(-C_A)$ }
\Line(107,20)(137,20)
\Text(230,20)[]{$=> ~~f^{cad}f^{dbc}=-C_A \delta^{ab}$ }
\end{picture}
\end{center}
\begin{equation}
\label{equation28}
\end{equation}
\vspace{-1.8cm}
\begin{center}
\begin{picture}(300,35)(0,0)
\Line(20,10)(60,10)
\Line(27,10)(40,30)
\Line(53,10)(40,30)
\Line(40,30)(40,35)

\Text(90,20)[]{$= - \frac{C_A}{2} $ }
\Line(120,35)(120,20)
\Line(120,20)(108,8)
\Line(120,20)(132,8)
\Text(230,20)[]{$=>~f^{dae}f^{ebg}f^{gcd}= - \frac12 C_A f^{abc} $ }
\end{picture}
\end{center}
\begin{center}
\begin{picture}(300,35)(0,0)
\Line(20,20)(60,40)
\Line(20,20)(60, 0)
\Line(7,20)(20,20)
\Line(40,30)(50,5)
\Line(40,10)(50,35)
\Text(265,18)[]{$.$}
\Text(80,20)[]{$=0$}
\Text(190,20)[]{$=>~f^{gai}f^{ijd}f^{jbh}f^{heg}f^{dce} = 0 $ }
\end{picture}
\end{center}
\vspace{-0.4cm}
The last two relations are derived from the Jacobi identity
%
%
%
%
\vspace{0.3cm}
\begin{equation}
\label{equation29}
\end{equation}
\vspace{-1.7cm}
\begin{center}
\begin{picture}(400,30)(0,0)
\Line(20,20)(45,20)
\Line(20,20)(15,35)
\Line(20,20)(15,5)
\Line(45,20)(50,35)
\Line(45,20)(50,5)

\Text(65,20)[]{$=$}
\Line(85,30)(85,10)
\Line(85,30)(73,35)
\Line(85,30)(97,35)

\Line(85,10)(73,5)
\Line(85,10)(97,5)
\Text(105,20)[]{$+$}

\Line(125,20)(150,20)
\Line(125,20)(120,5)
\Line(150,20)(155,5)
\Line(150,20)(120,35)
\Line(125,20)(155,35)
\Text(275,20)[]{$=>~~ f^{abc}f^{ade} + f^{abe}f^{acd} + f^{abd}f^{aec} = 0\,.$}
\end{picture}
\end{center}
The only products of structure constants which cannot be
contracted by the sequential use of (\ref{equation28})
are the following (Fig.6).
%
%
\begin{center}
\begin{picture}(200,50)(0,0)
\Line(20,0)(70,0)
\Line(45,40)(25,0)
\Line(45,40)(65,0)
\Line(45,40)(45,45)

\Line(45,0)(45,15)
\Line(35,20)(45,15)
\Line(55,20)(45,15)

\Vertex(45,40){1.0}
\Vertex(45,0){1.0}
\Vertex(25,0){1.0}
\Vertex(65,0){1.0}
\Vertex(45,15){1.0}
\Vertex(35,20){1.0}
\Vertex(55,20){1.0}

\Vertex(145,39){1.0}
\Vertex(125,0){1.0}
\Vertex(165,0){1.0}
\Vertex(153,0){1.0}
\Vertex(137,0){1.0}
\Vertex(139,27){1.0}
\Vertex(151,27){1.0}

\Line(120,0)(170,0)
\Line(145,40)(125,0)
\Line(145,40)(165,0)

\Line(145,40)(145,45)

\Line(139,27)(153,0)

\Line(151,27)(137,0)
\end{picture}
\end{center}
\begin{center}
Fig. 6
\end{center}
From (\ref{equation29}) we obtain
\begin{center}
\begin{picture}(200,50)(0,0)
\Line(20,0)(70,0)
\Line(45,40)(25,0)
\Line(45,40)(65,0)
\Line(45,40)(45,45)

\Line(45,0)(45,15)
\Line(35,20)(45,15)
\Line(55,20)(45,15)
\Text(100,15)[]{$= ~~-$}
\Text(190,15)[]{$- \frac18~ C_A^3$}

\Vertex(45,40){1.0}
\Vertex(45,0){1.0}
\Vertex(25,0){1.0}
\Vertex(65,0){1.0}
\Vertex(45,15){1.0}
\Vertex(35,20){1.0}
\Vertex(55,20){1.0}

\Line(120,0)(170,0)
\Line(145,40)(125,0)
\Line(145,40)(165,0)

\Line(145,40)(145,45)

\Line(139,27)(153,0)

\Line(151,27)(137,0)

\Line(230,20)(215,0)
\Line(230,20)(245,0)
\Line(230,20)(230,40)
\Vertex(230,20){1.0}

\Vertex(145,39){1.0}
\Vertex(125,0){1.0}
\Vertex(165,0){1.0}
\Vertex(153,0){1.0}
\Vertex(137,0){1.0}
\Vertex(139,27){1.0}
\Vertex(151,27){1.0}

\end{picture}
\end{center}
%
%
%
\vspace{-1.3cm}
\begin{equation}
\label{equation30}
\end{equation}
$$
$$
However, one fails to express the graphs of Fig.6 separately in terms
of $C_A$. In a specific case of the $SU(N)$ group, we have found
%
%
%
\begin{center}
\begin{picture}(170,50)(0,0)
\Text(90,15)[]{$= \frac32~ N$}
\Line(20,0)(70,0)
\Line(45,40)(25,0)
\Line(45,40)(65,0)

\Line(45,40)(45,45)

\Line(39,27)(53,0)

\Line(51,27)(37,0)

\Vertex(39,27){1.0}
\Vertex(53,0){1.0}
\Vertex(37,0){1.0}
\Vertex(51,27){1.0}
\Vertex(45,40){1.0}
\Vertex(25,0){1.0}
\Vertex(65,0){1.0}

\Line(130,20)(115,0)
\Line(130,20)(145,0)
\Line(130,20)(130,40)

\Vertex(130,20){1.0}

\end{picture}
\end{center}
\vspace{-1.3cm}
\begin{equation}
\label{equation31}
\end{equation}
$$
$$
Fortunately, the relation (\ref{equation30}) is quite sufficient for
the three-loop calculations of the renormalization group functions.
Only the sum of the diagrams of Fig.6 contributes to the final
answer. This fact is easy to explain. The non-trivial
products (Fig.6) might contribute to the vertex anomalous
dimension, $\tilde{\gamma}_1(\alpha, g^2)$, only. But it is known
to vanish in the Landau gauge: $\tilde{\gamma}_1(0, g^2)=0$.
Hence these products do not contribute to the gauge independent function
$\beta(g^2)$ and consequently, to $\tilde{\gamma}_1(\alpha, g^2)$
in arbitrary gauge as well.

Concluding this section we wish to discuss one more example
where Slavnov-Taylor identities [10] have been
fruitfully used. To facilitate the computation of the vertex
diagram with the two-loop three-gluon insertion
%
%
%
\begin{center}
\begin{picture}(300,50)(0,0)
\Photon(30,30)(30,40){2}{2}
\CArc(30,20)(10,0,180)
\CArc(30,20)(10,180,360)
\Text(30,20)[]{$2$}
\Text(70,15)[]{$=~{\cal K} $}
\DashArrowLine(50,0)(10,0){3}
\Text(0,15)[]{${\cal K} R'$}
\Photon(23,12)(18,0){2}{2}
\Photon(37,12)(42,0){2}{2}
\Photon(265,30)(265,40){2}{2}
\Text(274,36)[]{$\mu$}
\CArc(265,20)(10,0,180)
\CArc(265,20)(10,180,360)
\Text(266,20)[]{$2$}
\Photon(258,12)(249,0){2}{2}
\Photon(272,12)(279,0){2}{2}
\Text(243,0)[]{$\nu$}
\ArrowLine(262,8)(254,0)
\Text(262,-3)[]{$p$}

\Text(288,11)[]{$p$}
\ArrowLine(283,3)(278,12)
\Text(287,-3)[]{$\mu$}
\Line(230,0)(230,30)
\Line(230,30)(235,30)
\Line(230,0)(235,0)
\Line(300,0)(300,30)
\Line(295,0)(300,0)
\Line(295,30)(300,30)

\Text(237,16)[]{$R$}
\end{picture}
\end{center}
\vspace{-1.3cm}
\begin{equation}
 ~~~~{\frac{1}{4-2\varepsilon}}
\int \frac{ dp \ p_{\nu} }{(2\pi)^4 p^4 (p^2+m^2)}
\label{equation32}
\end{equation}
$$
$$
we employ an identity
\begin{equation}
p_{\mu}\Gamma_{\rho \nu \mu}^{abc}(k,q,p)
=G(p^2)\left[M^{abc}_{\sigma \rho}(k,q,p)
{{\cal{D}}}^{-1}(q^2)(q^2g_{\sigma \nu }-q_{\sigma}q_{\nu})
+\begin{pmatrix}
b \leftrightarrow a\\
\rho \leftrightarrow \nu \\
q \leftrightarrow k
\end{pmatrix}
\right],
\label{equation33}
\end{equation}
where a notation is as follows:
%
%
\begin{center}
\begin{picture}(190,55)(0,0)
\Photon(20,35)(20,50){2}{2}
\CArc(20,25)(10,0,180)
\CArc(20,25)(10,180,360)
\Photon(11,20)(4,8){2}{2}
\Photon(29,20)(36,8){2}{2}
\Text(100,25)[]{$=~\Gamma^{abc}_{\rho \nu \mu }(k,q,p)$}
\ArrowLine(14,49)(14,37)
\ArrowLine(-1,10)(6,20)
\ArrowLine(41,10)(34,20)

\Line(18,15)(30,27)
\Line(14,17)(28,31)
\Line(11,20)(25,34)
\Line(10,25)(20,35)

\Text(8,43)[]{$k$}
\Text(1,21)[]{$q$}
\Text(41,21)[]{$p$}
\Text(27,51)[]{$a$}
\Text(5,3)[]{$b$}
\Text(35,2)[]{$c$}
\Text(13,10)[]{$\nu$}
\Text(26,9)[]{$\mu$}

\Text(27,41)[]{$\rho$}
\Text(141,21)[]{$,$}
\end{picture}
\end{center}
\begin{center}
\begin{picture}(190,55)(0,0)
\Photon(20,35)(20,50){2}{2}
\CArc(20,25)(10,0,180)
\CArc(20,25)(10,180,360)
\DashArrowLine(12,19)(1,7){3}
\Photon(50,17)(29,28){2}{4}
\DashArrowLine(57,16)(27,16){3}
\Text(112,25)[]{$=~q_{\sigma}{ M}^{abc}_{\sigma \rho}(k,q,p)$}
\ArrowLine(14,49)(14,37)
\ArrowLine(0,13)(7,20)
\ArrowLine(54,11)(40,11)

\Line(18,15)(30,27)
\Line(14,17)(28,31)
\Line(11,20)(25,34)
\Line(10,25)(20,35)

\Text(8,43)[]{$k$}
\Text(1,21)[]{$p$}
\Text(41,30)[]{$\sigma$}
\Text(27,51)[]{$a$}
\Text(5,3)[]{$c$}
\Text(58,23)[]{$b$}
\Text(57,6)[]{$q$}

\Text(27,41)[]{$\rho$}
\Text(159,21)[]{$,$}
\end{picture}
\end{center}
\begin{center}
\begin{picture}(190,55)(0,0)
\CArc(35,25)(10,0,180)
\CArc(35,25)(10,180,360)
\DashArrowLine(3,25)(25,25){3}
\DashArrowLine(45,25)(67,25){3}
\Text(112,25)[]{$= - i \delta^{ab}~\frac{1}{p^2} ~{ G}(p^2)$}

\ArrowLine(3,18)(15,18)
\Line(33,15)(45,27)
\Line(29,17)(43,31)
\Line(26,20)(40,34)
\Line(25,25)(35,35)
\Text(1,12)[]{$p$}
\Text(5,32)[]{$a$}
\Text(59,32)[]{$b$}

\Text(156,21)[]{$,$}
\end{picture}
\end{center}
\begin{center}
\begin{picture}(230,55)(0,0)
\CArc(35,25)(10,0,180)
\CArc(35,25)(10,180,360)
\Photon(3,25)(25,25){2}{3}
\Photon(45,25)(67,25){2}{3}

\ArrowLine(8,18)(21,18)
\Line(33,15)(45,27)
\Line(29,17)(43,31)
\Line(26,20)(40,34)
\Line(25,25)(35,35)
\Text(13,10)[]{$p$}
\Text(5,32)[]{$a$}
\Text(59,32)[]{$b$}

\Text(0,18)[]{$\mu$}
\Text(63,17)[]{$\nu$}
\end{picture}
\end{center}
\vspace{-1.9cm}
\begin{equation}
\label{equation34}
~~~~~~~~~~~~~~~~~~~~~~~~~~~~
=-i\frac{\delta^{ab}}{p^2}\left[(g_{\mu\nu} - \frac{p_\mu p_\nu}{p^2})
 {\cal D}(p^2) +\alpha\frac{p_\mu p_\nu}{p^2}\right]\,.
\end{equation}
$$
$$
In our case $k=0$ so that (\ref{equation33}) transforms into
\begin{equation}
p_{\mu}\Gamma_{\rho \nu \mu}^{abc}(0, -p,p) = G(p^2){{\cal{D}}}^{-1}(p^2)
(p^2g_{\nu \sigma} - p_{\nu}p_{\sigma})
M^{abc}_{\sigma \rho}(0,-p,p)\,.
\label{equation35}
\end{equation}
Identity (\ref{equation35}) allows us to calculate $M^{abc}_{\sigma \rho}$
rather than fairly complicated three-gluon vertex
$\Gamma_{\rho \nu \mu}^{abc}$\,.

\vspace{.3cm}

\begin{center}
\large\textbf{5. \,Three-loop results for QCD}
\end{center}

A total number of topologically distinct three-loop diagrams
contributing to $\beta(g^2)$ amounts to 440 (without counting
opposite directions of the ghost and fermion lines).
For performing the Lorentz and Dirac algebra, reducing the integrands,
decomposing the scalar products, evaluating and summing
standard integrals, the computer program SCHOONSCHIP
[19] has been substantially used. The total execution
time is rather difficult to estimate. Here we only indicate that
the diagrams of Fig.7 require 110 and 90 seconds, respectively, at
the CDC-6500 computer.
%
%
%
%
\begin{center}
\begin{picture}(270,100)(0,0)
\PhotonArc(40,50)(30,0,180){2}{13}
\PhotonArc(40,50)(30,180,360){2}{13}

\Photon(18,69)(40,50){2}{4}
\Photon(40,50)(63,69){2}{4}
\Photon(40,50)(40,20){2}{4}

\Photon(-10,50)(10,50){2}{3}

\Photon(70,50)(90,50){2}{3}

\PhotonArc(230,50)(30,0,180){2}{12}
\PhotonArc(230,50)(30,180,360){2}{12}


\Photon(219,22)(219,77){2}{9}
\Photon(241,77)(241,22){2}{9}

\Photon(180,50)(200,50){2}{3}
\Photon(260,50)(280,50){2}{3}
\Text(132,-5)[]{${\rm~~Fig.~7}$ }
\end{picture}
\end{center}


Our final results obtained in collaboration with A.Yu. Zharkov
are as folows ($f$ is the number of flavors, $h=\frac{g^2}{(4\pi)^2}$):
\begin{eqnarray}
\label{equation36}
&&\tilde{\gamma}_1(1,h)= - \frac{C_A}{2}h - \frac34 C_A^2h^2
+h^3\left(- \frac{125}{32}C_A^3 + \frac{15}{8}C_A^2 T f \right),
\\
&& \nonumber \\
&& \gamma_3(1,h)= h\left(\frac53 C_A - \frac43 T f \right)
+h^2\left(\frac{23}{4}C_A^2 -5C_A T f -4 C_F T f \right)
\nonumber \\
&&    ~~~~~+ h^3\left[ \left(\frac{4051}{144}-\frac32 \zeta(3)  \right)
C_A^3 + \left( - \frac{875}{18} +18 \zeta(3) \right)C_A^2T f
\right.
\nonumber \\
&& ~~~~ \left.
-\left( \frac{5}{18} +24 \zeta(3) \right) C_AC_F T f
+ 2C_F^2 T f + \frac{76}{9}C_A T^2 f^2 + \frac{44}{9}C_F T^2 f^2
\right],
\label{equation37}
\end{eqnarray}
\begin{eqnarray}
&& \tilde{\gamma}_3(1,h)= \frac{C_A}{2}h + h^2 \left(
\frac{49}{24}C_A^2 - \frac56 C_A T f \right)
+ h^3 \left[ \left(\frac{229}{27}+ \frac34 \zeta(3) \right)
C_A^3
\right.
\nonumber \\
&& ~~~~\left. - \left(\frac{5}{216}+9 \zeta(3)\right)
C_A^2T f + \left(- \frac{45}{4}+12 \zeta(3) \right)
C_AC_F Tf - \frac{35}{27}C_AT^2 f^2  \right],
\label{equation38}
\end{eqnarray}
\begin{eqnarray}
&&\beta(h)=h^2\left(-\frac{11}{3}C_A + \frac43 Tf\right)
 +h^3\left(-\frac{34}{3}C_A^2 + \frac{20}{3} C_A T f + 4C_FT f \right)
\nonumber
\\
&&~~~~~~~~~~~~~~~~~~~~
+h^4\left(-\frac{2857}{54}C_A^3 + \frac{1415}{27} C_A^2Tf -\frac{158}{27}
C_AT^2f^2
\right.
\nonumber
\\
&&~~~~~~~~~~~~~~~~~~~~~~~~~~~
\left.
+ \frac{205}{9}C_AC_F Tf - \frac{44}{9}C_F T^2 f^2 -2C_F^2 T f \right).
\label{equation39}
\end{eqnarray}
The cancellation of the transcendental $\zeta(3)$ in the expression
for $\beta(h)$ is in complete analogy with QED treated in the
minimal subtraction scheme, where [20]
\begin{equation}
\beta_{QED}(\alpha) = \frac43 \frac{\alpha^2}{4\pi}
+ 4 \frac{\alpha^3}{(4\pi)^2}-\frac{62}{9} \frac{\alpha^4}{(4\pi)^3}.
\label{equation40}
\end{equation}
In a particular case of QCD, when fermions transform according
to the fundamental representation of $SU(3)$, $\beta(h)$
reads:
\begin{equation}
\beta_{QCD}(h) = h^2\!\left(\!-11+\frac23 f \!\right)
+ h^3 \!\left(\!-102 + \frac{38}{3}f \!\right)
+h^4 \!\left(\!-\frac{2857}{2} + \frac{5033}{18}f -\frac{325}{54}f^2\!\right).
\label{equation41}
\end{equation}
Now we are in a position to find an effective charge
 $\bar{h}\left(\frac{Q^2}{\mu^2},h\right)$ from
\begin{equation}
\ln \frac{Q^2}{\mu^2}= \int^{\bar{h}}_h \frac{dx}{\beta(x)}
= \psi(\bar{h})- \psi(h),
\label{equation42}
\end{equation}
where $\psi(h)$ represents an indefinite integral $\int^h\frac{dx}{\beta(x)}$.
Let us express $\bar{h}$ in  terms of renormalization group invariant
quantity $\ln \frac{Q^2}{\mu^2}+ \psi(h)\equiv \ln
\frac{Q^2}{\Lambda^2}\equiv L$, where $\Lambda$ is the momentum scale.
Assuming
\begin{equation}
\beta(x)= - \beta_0 x^2 -\beta_1 x^3 -\beta_2 x^4 + O(x^5)
\label{equation43}
\end{equation}
we arrive at
\begin{equation}
\psi(h) = \frac{1}{\beta_0 h} + \frac{\beta_1}{\beta_0^2} \ln h + \delta
+ \frac{\beta_2 \beta_0 - \beta_1^2}{\beta_0^3 }h + O(h^2)
\label{equation44}
\end{equation}
and obtain from (\ref{equation42})
\begin{multline}
\label{equation45}
\bar{h}(L)=\frac{1}{\beta_0L }- \frac{\beta_1}{\beta_0^3}
\frac{\ln L}{L^2} + \frac{\delta \beta_0^2 -\beta_1 \ln \beta_0}
{\beta_0^3 L^2}+ \frac{\beta_1^2\ln^2 L}{\beta_0^5 L^3}
- \frac{\ln L}{L^3}\left[\frac{\beta_1^2}{\beta_0^5} +
\frac{2\beta_1}{\beta_0^5}\left(\delta \beta_0^2 - \beta_1 \ln \beta_0
\right) \right] \\
+ \frac{1}{L^3\beta_0^5}\left[\beta_2 \beta_0 -\beta_1^2
 +\beta_1 (\delta \beta_0^2 - \beta_1 \ln \beta_0)
+ (\delta \beta_0^2 - \beta_1 \ln \beta_0)^2  \right]
+ O\left( \frac{\ln^3L}{L^4}\right)
\end{multline}
with $\delta$ being an arbitrary constant. Fixing the momentum scale
$\Lambda$ by choosing, as usual,
$\delta = \frac{\beta_1\ln \beta_0}{\beta_0^2}$, we finally get
\begin{equation}
\bar{h}(L)= \frac{1}{\beta_0 L}
- \frac{\beta_1}{\beta_0^3}\frac{\ln L}{L^2}
+ \frac{\beta_1^2 (\ln^2 L - \ln L)}{\beta_0^5 L^3}
+ \frac{\beta_2\beta_0 - \beta_1^2}{\beta_0^5 L^3}
+O\left(\frac{\ln^3 L}{L^4}\right).
\label{equation46}
\end{equation}
Using (\ref{equation41}), (\ref{equation43}) and (\ref{equation46})
one readily finds the QCD effective charge in the three-loop approximation.

\vspace{.3cm}

\begin{center}
\large\textbf{6. \,Vanishing of $\beta(g^2)$ in a supersymmetric gauge model}
\end{center}

Some time ago a very interesting $SU(4)$-supersymmetric
non-Abelian gauge model has been derived [3,4]
which exhibits the vanishing charge renormalization effects, since its charge
renormalization function $\beta(g^2)$ proves to be zero through the
two-loop order [5]. The Lagrangian is [4]:
$$
{\cal L} = {{\cal{L}_{\rm YM}}} + \frac{i}{2}\bar{\lambda}^a_m
{\hat{\cal{D}}}\lambda^a_m
+ \frac12 \left( {{\cal{D}_{\mu}}} \phi_r^a \right)^2
+ \frac12 \left( {{\cal{D}_{\mu}}} \chi_r^a \right)^2
-\frac{g}{2}f^{abc}\bar{\lambda}^a_m
\left[ \alpha^r_{mn}\phi^b_r + \gamma_5\beta^r_{mn}
\chi^b_r \right]\lambda^c_n
$$
\begin{equation}
\label{equation47}
-\frac{g^2}{4}\left[ (f^{abc}\phi^b_r \phi^c_t)^2
+(f^{abc}\chi^b_r \chi^c_t)^2
+2(f^{abc}\phi^b_r \chi^c_t)^2 \right],
\end{equation}
with $a,b,c=1,...,N^2-1;~m,n=1,...,4;~r,t=1,2,3$. Here
$ {{\cal{L}_{\rm YM}}}$ is the pure Yang-Mills Lagrangian with $SU(N)$ gauge symmetry.
The matter fields (Majorana spinors $\lambda_m^a$, scalars
$\phi^a_r$ and pseudoscalars $\chi^a_r$) transform according to the adjoint
(regular) representation of $SU(N)$. Hence
$$
{{\cal{D}_{\mu}}}\lambda^a_m= \partial_{\mu}\lambda^a_m
+ g f^{abc} A_{\mu}^b \lambda^c_m
$$
with similar expressions for ${{\cal{D}_{\mu}}}\phi^a_r$ and
 ${{\cal{D}_{\mu}}}\chi^a_r$. The six real
antisymmetric $4\times 4$ matrices $\alpha^r$, $\beta^r$ obey the relations
\begin{equation}
[\alpha^r, \alpha^t]_{+}= [\beta^r, \beta^t]_{+}= -2\delta^{rt},~~~~
[\alpha^r,\beta^t]_{-}=0.
\label{equation48}
\end{equation}
The other properties of these matrices and their explicit form
are given in Appendix.

To determine the contributions to the renormalization group
functions of the model (\ref{equation47}) from the diagrams without
scalar and pseudoscalar particles, one may use the results
(\ref{equation36})-(\ref{equation39}) with
\begin{equation}
C_A=C_F=N, ~~~~Tf=2N.
\label{equation49}
\end{equation}
This leads to
\begin{equation}
\beta(h)_{\rm without ~scalars}=-Nh^2 + 10N^2h^3 + \frac{101}{2}N^3h^4.
\label{equation50}
\end{equation}
Now an appropriate scalar contribution must be added to (\ref{equation50}).
In the two-loop approximation it has been done in [5]
with the intriguing result $\beta(h)=0$.

The method of our three-loop calculations is described above.
Here we shall only consider the issue of applicability of the standard
dimensional regularization to supersymmetric
theories. This subject has been discussed by various authors [21].
Proceeding in the spirit of Ref.~[21] we write
down the following rules of the ``supersymmetric dimensional
regularization" which is to maintain both gauge invariance and
global supersymmetry:
The relations defining the Dirac matrices look as in four dimensions
(see Appendix) while the numbers of scalar and pseudoscalar fields equal
$3+\varepsilon$ rather than 3. This modification of the regularization
maintains equal (and integral) total numbers of Bose and Fermi
degrees of freedom
even in $4-2\varepsilon$ dimensions: 8 components of four Majorana spinors
correspond to $(2-2\varepsilon)$ massless vectors + $(3+\varepsilon)$
scalars + $(3+\varepsilon)$ pseudoscalars $=8$ bosons. It is this
matching of the Fermi and Bose field components that is crucial
for preserving supersymmetry [21].

For lack of a rigorous proof, we have verified the invariance
of the supersymmetric dimensional regularization by
direct calculation of $\beta(h)$ at the two-loop level in two
different ways:
\begin{equation}
\beta(h)=h[2\widetilde{\gamma}_1(h) -\gamma_3(h)
-2\widetilde{\gamma}_3(h)]
\label{equation51}
\end{equation}
and
\begin{equation}
\beta(h) = h[2\gamma_4(h)-\gamma_{\phi}(h)
-2\gamma_{\lambda}(h)].
\label{equation52}
\end{equation}
Here $\widetilde{\gamma}_1$ and $\gamma_4$ are the anomalous dimensions
of the ghost-ghost-gluon and fermion-fermion-scalar vertices,
and $\gamma_3$, $\widetilde{\gamma}_3$,
${\gamma}_{\phi}$ and ${\gamma}_{\lambda}$ are those of gluon,
ghost, scalar and fermion propagators, respectively. In the standard
(with $\delta^{rr}=3$) dimensional regularization, these anomalous
dimensions are (in the Feynman gauge):
\begin{eqnarray}
&&
\widetilde{\gamma}_1 = -\frac{Nh}{2} - \frac{3}{4}N^2h^2,
~~~~~~~\gamma_4=-5Nh + 5N^2 h^2,
\nonumber
\\
\label{equation53}
&&\gamma_3=-2Nh + \frac{N^2h^2}{2}, ~~~~~~~~
\gamma_{\phi}=-2Nh,
\\
&&\widetilde{\gamma}_3=\frac{Nh}{2}-N^2h^2,~~~~~~~~~~~~
\gamma_{\lambda}=-4Nh+6N^2h^2.
\nonumber
\end{eqnarray}
With the use of supersymmetric dimensional regularization
(with $\delta^{rr}=3+\varepsilon$), we obtain
\begin{eqnarray}
&&
\widetilde{\gamma}_1 = -\frac{Nh}{2} - \frac{3}{4}N^2h^2,
~~~~~~~~~\gamma_4=-5Nh + \frac{11}{2}N^2 h^2,
\nonumber
\\
\label{equation54}
&&\gamma_3=-2Nh + N^2h^2, ~~~~~~~~~~
\gamma_{\phi}=-2Nh-N^2h^2,
\\
&&\widetilde{\gamma}_3=\frac{Nh}{2}-\frac{5}{4}N^2h^2,~~~~~~~~~~~~
\gamma_{\lambda}=-4Nh+6N^2h^2.
\nonumber
\end{eqnarray}
Using (\ref{equation51}) gives $\beta(h)=0$ for both regularizations
while (\ref{equation52}) leads to $\beta(h)=-2N^2h^3$ for the
standard regularization and to $\beta(h)=0$ for the supersymmetric
one. This discrepancy shows the former regularization to be
noninvariant under supersymmetric transformations.

For our three-loop calculations we employ formula (\ref{equation51}).
Below we write down the scalar contributions to anomalous
dimensions through the three-loop order calculated in the
supersymmetric dimensional regularization scheme (in collaboration
with L.V. Avdeev):
\begin{eqnarray}
&&\gamma^{scal}_3 =\,-Nh+\frac{53}{4}N^2h^2 +\left(\frac{69}{8}
-\frac{9}{4}\,\zeta(3) \right) N^3h^3,
\nonumber
\\
\label{equation55}
&&\widetilde{\gamma}^{scal}_3 = ~~~~~~~~-\frac{13}{8}N^2h^2
+\left(\frac{771}{32}+\frac{9}{8}\,\zeta(3) \right)N^3h^3,
\\
&&
\widetilde{\gamma}_1^{scal} = ~~~~~~~~~~~~~~~~~~~~~~~~~~~\frac{101}{32}N^3h^3.
\nonumber
\end{eqnarray}
From (\ref{equation55}) and (\ref{equation51}) we obtain
\begin{equation}
\beta^{scal}(h)=Nh^2-10N^2h^3 -\frac{101}{2} N^3h^4
\label{equation56}
\end{equation}
and using (\ref{equation50}), arrive at the final result
\begin{equation}
\beta(h)_{\rm three~loops}=0.
\label{equation57}
\end{equation}
It is worth mentioning that the use of the standard dimensional
regularization yields
\begin{eqnarray}
&&\gamma_3^{scal}=-Nh + \frac{51}{4} N^2h^2
+\left(\frac{193}{48} -\frac{9}{4}\,\zeta(3) \right) N^3h^3,
\nonumber \\
&&\widetilde{\gamma}_3^{scal}= -\frac{11}{8}N^2h^2
+\left(\frac{527}{24}+\frac{9}{8}\,\zeta(3)\right)N^3h^3,~~~~~
\widetilde{\gamma}_1^{scal}=\frac{87}{32}N^3h^3,
\\
&&\beta(h)_{three~loops}= 8N^3h^4.
\nonumber
\label{equation58}
\end{eqnarray}

The result (\ref{equation57}) implies the absence of the
charge renormalization effects in the model (\ref{equation47})
to the three-loop order.
It confirms a conjecture that $\beta(h)$ in this model vanishes
to all orders. If it were the case, the model (\ref{equation47})
would be unique in the four dimensional quantum field theory.
The vanishing $\beta(h)$ might imply, for instance, that this
model would be free of supersymmetric anomalies [22].
In any case, a rigorous argument proving this
conjecture on symmetry ground is now a great urgency.

\vspace{.1cm}

We would like to thank L.V.\,Avdeev, G.A.\,Chochia and
A.Yu.\,Zharkov for the help in some calculations.

\vspace{.2cm}

\begin{center}
\textbf{APPENDIX}
\end{center}

\noindent
\textbf{Feynman rules for the model (1)}
\begin{center}
\begin{picture}(300,40)(0,0)
\Photon(20,20)(50,20){2}{5}
\Text(10,20)[]{$A_{\mu}^a$}
\Text(60,20)[]{$A_{\nu}^b$ }
\Text(120,20)[]{$-$}
\Text(134,27)[]{$i$}
\Line(130,20)(140,20)
\Text(134,12)[]{$p^2$}
\Text(150,20)[]{$\delta^{ab}$}
\Text(160,20)[]{$\Big ($}
\Text(200,20)[]{$g_{\mu \nu} + (\alpha-1)$}
\Text(265,20)[]{$\Big )\ ,$}
\Line(235,20)(255,20)
\Text(247,27)[]{$p_{\mu}p_{\nu}$}
\Text(244,12)[]{$p^2$}

\end{picture}
\end{center}

\begin{center}
\begin{picture}(300,40)(0,0)
\DashArrowLine(20,20)(50,20){3}
\Text(10,20)[]{$\overline{\eta}^a$}
\Text(60,21)[]{$\eta^b$ }

\Text(120,20)[]{$-$}
\Text(134,27)[]{$i$}
\Line(130,20)(140,20)
\Text(134,12)[]{$p^2$}
\Text(155,20)[]{$\delta^{ab}\ ,$}
\end{picture}
\end{center}
\begin{center}
\begin{picture}(300,40)(0,0)
\ArrowLine(20,20)(50,20)
\Text(10,20)[]{$\overline{\psi}^{~m}_i$}
\Text(60,21)[]{$\psi^n_j$ }
\ArrowLine(30,26)(40,26)
\Text(35,33)[]{$p$}
\Text(134,27)[]{$i \hat{p}$}
\Line(130,20)(140,20)
\Text(134,12)[]{$p^2$}
\Text(163,20)[]{$\delta^{mn}\delta_{ij}\ ,$}
\end{picture}
\end{center}
\begin{center}
\begin{picture}(300,40)(0,0)
\Photon(10,20)(40,20){2}{3}

\Text(44,40)[]{$p$}
\ArrowLine(53,39)(44,30)
\DashArrowLine(55,35)(40,20){2}
\DashArrowLine(40,20)(55,5){2}
\Text(10,26)[]{$\mu$}
\Text(10,15)[]{$c$}
\Text(60,35)[]{$a$}
\Text(60,5)[]{$b$}

\Text(147,20)[]{$ g\,p_\mu\,f^{abc}~,$}
\end{picture}
\end{center}

\begin{center}
\begin{picture}(300,40)(0,0)
\Photon(10,20)(40,20){2}{3}
\ArrowLine(55,35)(40,20)
\ArrowLine(40,20)(55,5)
\Text(10,26)[]{$\mu$}
\Text(10,15)[]{$a$}

\Text(60,35)[]{$j$}
\Text(60,5)[]{$i$}
\Text(48,38)[]{$n$}
\Text(48,3)[]{$m$}

\Text(149,20)[]{$ig\gamma_\mu\delta^{mn}R^a_{ij}\ ,$}
\end{picture}
\end{center}
\begin{center}
\begin{picture}(320,50)(0,0)
\Photon(10,20)(40,20){2}{3}
\Photon(55,35)(40,20){2}{3}
\Photon(40,20)(55,5){2}{3}

\Text(10,27)[]{$\alpha$}
\Text(10,15)[]{$a$}

\Text(60,40)[]{$b$}
\Text(60,0)[]{$c$}
\Text(48,38)[]{$\beta$}
\Text(47,3)[]{$\gamma$}

\Text(223,20)[]{$g\,f^{abc}\,[(p-q)_\alpha\,g_{\beta\gamma} + (q-k)_\beta\,g_{\alpha\gamma}
 + (k-p)_\gamma\,g_{\alpha\beta}]$}
\Text(348,18)[]{$,$}
\Text(25,34)[]{$k$}
\ArrowLine(19,26)(31,26)
\ArrowLine(59,32)(49,22)
\ArrowLine(59,8)(49,18)
\Text(63,29)[]{$p$}
\Text(63,10)[]{$q$}
\end{picture}
\end{center}
\begin{center}
\begin{picture}(340,50)(0,0)
\Photon(25,40)(58,5){2}{6}
\Photon(25,5)(58,40){2}{6}
\Text(20,5)[]{$a$}
\Text(63,5)[]{$d$}
\Text(20,40)[]{$b$}
\Text(63,40)[]{$c$}

\Text(56,44)[]{$\mu$}
\Text(34,44)[]{$\beta$}
\Text(32,1)[]{$\alpha$}
\Text(52,1)[]{$\nu$}
\Text(223,40)[]{$
-ig^2\,[f^{abe}f^{cde}\,(2g_{\alpha\mu}g_{\beta\nu}
 - g_{\alpha\nu}g_{\beta\mu} - g_{\alpha\beta}g_{\mu\nu})
 $}
\Text(223,10)[]{$ ~~~~+ f^{ace}f^{bde}\,(2g_{\alpha\beta}g_{\mu\nu}
 - g_{\alpha\nu}g_{\beta\mu} - g_{\alpha\mu}g_{\beta\nu})]
 $}
\Text(335,10)[]{$.$}
\end{picture}
\end{center}

\vspace{.5cm}

\noindent
\textbf{Additional Feynman rules for the model (47)}

\vspace{.0cm}
\begin{center}
\begin{picture}(300,50)(0,0)
\Line(10,15)(45,15)
\ArrowLine(23,20)(35,20)
\Text(29,28)[]{$p$}
\Text(0,16)[]{$\overline{\lambda}^a_m$}
\Text(54,16)[]{$\lambda^b_n$}
\Text(150,25)[]{$i\hat{p}$}
\Text(150,7)[]{$p^2$}
\Text(180,15)[]{$\delta^{ab}\delta_{mn}\ ,$}
\Line(145,15)(155,15)

\end{picture}
\end{center}
\begin{center}
\begin{picture}(300,40)(0,0)
\Vertex(15,20){0.8}
\Vertex(20,20){0.8}
\Vertex(25,20){0.8}
\Vertex(30,20){0.8}
\Vertex(35,20){0.8}
\Vertex(40,20){0.8}
\Vertex(45,20){0.8}
\Vertex(50,20){0.8}
\Vertex(55,20){0.8}
\Text(66,20)[]{$\phi^b_t$}
\Text(2,20)[]{$\phi^a_r$}

\Line(15,5)(21,5)
\Vertex(24,5){0.7}
\Line(27,5)(33,5)
\Vertex(36,5){0.7}
\Line(39,5)(45,5)
\Vertex(48,5){0.7}
\Line(51,5)(57,5)
\Text(66,5)[]{$\chi^b_t$}
\Text(2,5)[]{$\chi^a_r$}
\Text(89,12)[]{$\Big \}$}

\Text(150,23)[]{$i$}
\Text(150,7)[]{$p^2$}
\Text(180,15)[]{$\delta^{ab}\delta_{rt}\ ,$}
\Line(145,15)(153,15)
\end{picture}
\end{center}

\begin{center}
\begin{picture}(300,40)(0,0)
\Photon(10,20)(40,20){2}{3}

\ArrowLine(55,35)(40,20)
\ArrowLine(40,20)(55,5)
\Text(10,26)[]{$\mu$}
\Text(10,15)[]{$c$}

\Text(60,35)[]{$a$}
\Text(60,5)[]{$b$}
\Text(48,38)[]{$m$}
\Text(48,3)[]{$n$}

\Text(159,20)[]{$-~g\gamma_\mu f^{abc} \delta_{mn}\ , $}
\end{picture}
\end{center}

\begin{center}
\begin{picture}(300,45)(0,0)
\Text(40,36)[]{$k$}
\ArrowLine(49,35)(39,25)
\ArrowLine(39,14)(49,4)
\Text(40,4)[]{$p$}
\Photon(10,20)(40,20){2}{3}
\Vertex(43,23){0.7}
\Vertex(46,26){0.7}
\Vertex(49,29){0.7}
\Vertex(52,32){0.7}
\Vertex(55,35){0.7}
\Vertex(43,17){0.7}
\Vertex(46,14){0.7}
\Vertex(49,11){0.7}
\Vertex(52,8){0.7}
\Vertex(55,5){0.7}

\Text(10,26)[]{$\mu$}
\Text(10,15)[]{$c$}

\Text(62,35)[]{$a$}
\Text(62,5)[]{$b$}
\Text(54,40)[]{$r$}
\Text(53,0)[]{$t$}

%
\Text(130,36)[]{$k$}
\ArrowLine(139,35)(129,25)
\ArrowLine(129,14)(139,4)
\Text(130,4)[]{$p$}

\Photon(100,20)(130,20){2}{3}
\Line(131,21)(135,25)
\Vertex(137,27){0.7}
\Line(139,29)(143,33)
\Vertex(145,35){0.7}

\Line(131,19)(135,15)
\Vertex(137,13){0.7}
\Line(139,11)(143,7)
\Vertex(145,5){0.7}

\Text(100,26)[]{$\mu$}
\Text(100,15)[]{$c$}

\Text(152,35)[]{$a$}
\Text(152,5)[]{$b$}
\Text(145,42)[]{$r$}
\Text(142,0)[]{$t$}
\Text(80,20)[]{$=$}
\Text(240,20)[]{$-~g (k+p)_{\mu}~ f^{abc} \delta_{rt}\ , $}
\end{picture}
\end{center}
\begin{center}
\begin{picture}(300,40)(0,0)

\Vertex(10,20){0.7}
\Vertex(15,20){0.7}
\Vertex(20,20){0.7}
\Vertex(25,20){0.7}
\Vertex(30,20){0.7}
\Vertex(35,20){0.7}
\ArrowLine(55,35)(40,20)
\ArrowLine(40,20)(55,5)
\Text(10,26)[]{$r$}
\Text(10,15)[]{$c$}

\Text(60,35)[]{$a$}
\Text(60,5)[]{$b$}
\Text(48,38)[]{$m$}
\Text(48,3)[]{$n$}

\Text(159,20)[]{$-~i g f^{abc} ~\alpha_{nm}^r\ , $}
\end{picture}
\end{center}

\begin{center}
\begin{picture}(300,40)(0,0)

\Line(10,20)(15,20)
\Vertex(17,20){0.7}
\Line(19,20)(24,20)
\Vertex(26,20){0.7}
\Line(28,20)(33,20)
\Vertex(35,20){0.7}
\Line(37,20)(40,20)

\ArrowLine(55,35)(40,20)
\ArrowLine(40,20)(55,5)
\Text(10,26)[]{$r$}
\Text(10,15)[]{$c$}

\Text(60,35)[]{$a$}
\Text(60,5)[]{$b$}
\Text(48,38)[]{$m$}
\Text(48,3)[]{$n$}

\Text(159,20)[]{$-~i g f^{abc} ~\gamma_5~\beta_{nm}^r\ , $}
\end{picture}
\end{center}

\begin{center}
\begin{picture}(340,50)(0,0)

\Text(18,41)[]{$\mu$}
\Text(28,46)[]{$a$}
\Text(18,5)[]{$\nu$}
\Text(28,0)[]{$b$}

\Text(65,41)[]{$r$}
\Text(56,46)[]{$c$}
\Text(65,5)[]{$t$}
\Text(55,0)[]{$d$}
\Photon(24,41)(42,23){2}{4}
\Photon(24,5)(42,23){2}{4}

\Vertex(42,23){0.7}
\Vertex(45,26){0.7}
\Vertex(48,29){0.7}
\Vertex(51,32){0.7}
\Vertex(54,35){0.7}
\Vertex(57,38){0.7}
\Vertex(60,41){0.7}

\Vertex(45,20){0.7}
\Vertex(48,17){0.7}
\Vertex(51,14){0.7}
\Vertex(54,11){0.7}
\Vertex(57,8){0.7}
\Vertex(60,5){0.7}

\Text(80,23)[]{$=$}
\Text(98,41)[]{$\mu$}
\Text(108,46)[]{$a$}
\Text(98,5)[]{$\nu$}
\Text(108,0)[]{$b$}

\Text(145,38)[]{$r$}
\Text(136,46)[]{$c$}
\Text(148,5)[]{$t$}
\Text(135,0)[]{$d$}
\Photon(104,41)(122,23){2}{4}
\Photon(104,5)(122,23){2}{4}

\Vertex(122,23){0.7}
\Line(124,25)(128,29)
\Vertex(130,31){0.7}
\Line(132,33)(136,37)
\Vertex(138,39){0.7}
\Line(140,41)(144,45)

\Line(124,21)(128,17)
\Vertex(130,15){0.7}
\Line(132,13)(136,9)
\Vertex(138,7){0.7}
\Line(140,5)(144,1)

\Text(260,23)[]{$ig^2\,g_{\mu\nu}\delta_{rt}(f^{ace}f^{bde} + f^{ade}f^{bce})\ ,$}
\end{picture}
\end{center}

\begin{center}
\begin{picture}(340,50)(0,0)

\Text(58,41)[]{$r$}
\Text(68,46)[]{$a$}
\Text(58,5)[]{$t$}
\Text(68,0)[]{$b$}

\Text(105,38)[]{$s$}
\Text(96,46)[]{$c$}
\Text(108,5)[]{$u$}
\Text(95,0)[]{$d$}
\Vertex(82,23){0.7}
\Vertex(79,26){0.7}
\Vertex(76,29){0.7}
\Vertex(73,32){0.7}
\Vertex(70,35){0.7}
\Vertex(67,38){0.7}
\Vertex(64,41){0.7}

\Vertex(79,20){0.7}
\Vertex(76,17){0.7}
\Vertex(73,14){0.7}
\Vertex(70,11){0.7}
\Vertex(67,8){0.7}
\Vertex(64,5){0.7}

\Line(84,25)(88,29)
\Vertex(90,31){0.7}
\Line(92,33)(96,37)
\Vertex(98,39){0.7}
\Line(100,41)(104,45)

\Line(84,21)(88,17)
\Vertex(90,15){0.7}
\Line(92,13)(96,9)
\Vertex(98,7){0.7}
\Line(100,5)(104,1)

\Text(260,23)[]{$-ig^2\,\delta_{rt}\delta_{su}(f^{ace}f^{bde} + f^{ade}f^{bce})\ ,$}
\end{picture}
\end{center}

\begin{center}
\begin{picture}(340,50)(0,0)

\Text(18,41)[]{$r$}
\Text(28,46)[]{$a$}
\Text(18,5)[]{$s$}
\Text(28,0)[]{$b$}

\Text(65,41)[]{$t$}
\Text(56,46)[]{$c$}
\Text(65,5)[]{$u$}
\Text(55,0)[]{$d$}

\Vertex(39,26){0.7}
\Vertex(36,29){0.7}
\Vertex(33,32){0.7}
\Vertex(30,35){0.7}
\Vertex(27,38){0.7}
\Vertex(24,41){0.7}

\Vertex(39,20){0.7}
\Vertex(36,17){0.7}
\Vertex(33,14){0.7}
\Vertex(30,11){0.7}
\Vertex(27,8){0.7}
\Vertex(24,5){0.7}

\Vertex(42,23){0.7}
\Vertex(45,26){0.7}
\Vertex(48,29){0.7}
\Vertex(51,32){0.7}
\Vertex(54,35){0.7}
\Vertex(57,38){0.7}
\Vertex(60,41){0.7}

\Vertex(45,20){0.7}
\Vertex(48,17){0.7}
\Vertex(51,14){0.7}
\Vertex(54,11){0.7}
\Vertex(57,8){0.7}
\Vertex(60,5){0.7}

\Text(80,23)[]{$=$}
\Text(96,41)[]{$r$}
\Text(108,46)[]{$a$}
\Text(96,5)[]{$s$}
\Text(108,0)[]{$b$}

\Text(145,38)[]{$t$}
\Text(136,46)[]{$c$}
\Text(148,5)[]{$u$}
\Text(135,0)[]{$d$}

\Line(120,25)(116,29)
\Vertex(114,31){0.7}
\Line(112,33)(108,37)
\Vertex(106,39){0.7}
\Line(104,41)(100,45)

\Line(120,21)(116,17)
\Vertex(114,15){0.7}
\Line(112,13)(108,9)
\Vertex(106,7){0.7}
\Line(104,5)(100,1)

\Vertex(122,23){0.7}
\Line(124,25)(128,29)
\Vertex(130,31){0.7}
\Line(132,33)(136,37)
\Vertex(138,39){0.7}
\Line(140,41)(144,45)

\Line(124,21)(128,17)
\Vertex(130,15){0.7}
\Line(132,13)(136,9)
\Vertex(138,7){0.7}
\Line(140,5)(144,1)
\Text(270,38)[]{$-ig^2\,[f^{abe}f^{cde}(2\delta_{rt}\delta_{su}
 - \delta_{rs}\delta_{tu} - \delta_{ru}\delta_{ts})$}
\Text(270,4)[]{$ ~~~~+ f^{ace}f^{bde}(2\delta_{rs}\delta_{tu}
 - \delta_{rt}\delta_{su} - \delta_{ru}\delta_{ts})]\ .$}
\end{picture}
\end{center}
In addition to this: \\
a) each closed loop brings a factor $(2\pi)^{-4}$, \\
b) each fermion or ghost loop gives an extra minus sign, \\
c) arrows on the Majorana spinor lines should be ignored in
calculating the symmetry factors.

\vspace{.5cm}

\noindent
\textbf{Dirac matrices in $4-2\varepsilon$ dimensions}

\vspace{.2cm}

We use the metric $g_{\mu\nu}=(1,-1,-1,...), \ g_{\mu \mu}=4-2\varepsilon$\,.
\begin{eqnarray}
&&
[\gamma_{\mu},\gamma_{\nu}]_+ = 2g_{\mu \nu},~~~~
\gamma_{\mu}\gamma_{\mu}=4-2\varepsilon,
~~~~
\gamma_{\mu}\gamma_{\nu} \gamma_{\mu} = (2\varepsilon-2)
\gamma_{\nu},
\nonumber \\
&&
\gamma_{\mu}\gamma_{\nu}\gamma_{\rho}\gamma_{\mu}=
4g_{\nu \rho} -2 \varepsilon \gamma_{\nu}\gamma_{\rho},
~~~\gamma_{\mu}\gamma_{\nu}\gamma_{\rho}\gamma_{\sigma}\gamma_{\mu}
 = 2 \varepsilon\,\gamma_{\nu}\gamma_{\rho}\gamma_{\sigma}
-2\gamma_{\sigma}\gamma_{\rho}\gamma_{\nu},
\\
&&
[\gamma_{5},\gamma_{\mu}]_{+} = 0, ~~\gamma_5^2=-1,~~
{\rm tr}\,\gamma_5=0, ~~{\rm tr}\,I =4,
~~{\rm tr}(\gamma_{\mu} \gamma_{\nu}) = 4 g_{\mu \nu},
\nonumber
\\
&&{\rm tr}(\gamma_{\mu}\gamma_{\nu}
\gamma_{\alpha}\gamma_{\beta})
=4(g_{\mu \nu}g_{\alpha \beta}-g_{\mu \alpha}
g_{\nu \beta} + g_{\mu \beta} g_{\nu \alpha}),
~~~{\rm tr}(\gamma_{\mu_1}... \gamma_{\mu_{2N+1}})=0\,.
\nonumber
\label{equationA1}
\end{eqnarray}

\vspace{.5cm}

\noindent
\textbf{The $\alpha$- and $\beta$-matrices of the model (47)}

\vspace{.2cm}

These real antisymmetric $4 \times 4$ matrices have an explicit
representation in terms of the Pauli matrices:
\begin{eqnarray}
&&\alpha^1=\left( \begin{array}{cc}
0         & \sigma_1\\
-\sigma_1 & 0
\end{array}
\right),~~
\alpha^2=\left( \begin{array}{cc}
0         & -\sigma_3\\
\sigma_3 & 0
\end{array}
\right),~~
\alpha^3=\left( \begin{array}{cc}
i \sigma_2         & 0 \\
0 & i \sigma_2
\end{array}
\right),
\nonumber \\
&& \label{equationA2} \\
&&\beta^1=\left( \begin{array}{cc}
0         & i \sigma_2\\
i\sigma_2 & 0
\end{array}
\right),~~
\beta^2=\left( \begin{array}{cc}
0         & 1\\
-1 & 0
\end{array}
\right),~~
\beta^3=\left( \begin{array}{cc}
-i \sigma_2         & 0 \\
0 & i \sigma_2
\end{array}
\right).
\nonumber
\end{eqnarray}
Their relevant properties are
\begin{eqnarray}
&&[\alpha^r,\alpha^t]_{+}=[\beta^r, \beta^t]_{+}
=-2\delta^{rt},~~~[\alpha^r,\beta^t]_{-}=0,
\nonumber \\
&& \label{equationA3} \\
&& {\rm tr}\,\alpha^r={\rm tr}\,\beta^r
= {\rm tr}\,(\alpha^r \beta^t) =0, ~~~
{\rm tr}(\alpha^r \alpha^t) =
{\rm tr}(\beta^r \beta^t) = -4 \delta^{rt}.
\nonumber
\end{eqnarray}
The supersymmetric regularization used in section 6
implies $\delta^{rr}=3+\varepsilon$ giving rise to the
following relations:
\begin{equation}
\alpha^r \alpha^r = \beta^r\beta^r = -3-\varepsilon,~~~
\alpha^r \alpha^t \alpha^r =(1+\varepsilon)\alpha^t,~~~
\beta^r \beta^t \beta^r =(1+\varepsilon)\beta^t,
\label{equationA4}
\end{equation}
whereas the standard dimensional regularization
prescribes
\begin{equation}
\delta^{rr}=3,~~~
\alpha^r \alpha^r = \beta^r\beta^r = -3,~~~
\alpha^r \alpha^t \alpha^r = \alpha^t,~~~
\beta^r \beta^t \beta^r = \beta^t\,.
\label{equationA5}
\end{equation}

\vspace{.5cm}

\noindent
\textbf{Properties of the Euler $\Gamma$-function}

\vspace{0cm}

\begin{eqnarray}
&&\Gamma(z+1)=z\Gamma(z),~~~~\Gamma(1)=\Gamma(2)=1, \ \ \ \
\Gamma(N+1)=N!,
\nonumber \\
&& \label{equationA6} \\
&&\Gamma(1+x)=\exp \left[ -\gamma x + \sum_{n=2}^{\infty}
(-1)^n\frac{\zeta(n)}{n}x^n \right],
\nonumber
\end{eqnarray}
where $\gamma$ is the Euler constant and $\zeta$ the Riemann
function.
We note that $\gamma$ and $\zeta(2)$ do not occur in
${{\cal{K}R'}G}$, and consequently in the renormalization
group functions.

\vspace{.5cm}

\noindent
\textbf{One-loop integration formulas}

\vspace{.2cm}

We choose a volume of the unit sphere in $4-2\varepsilon$
dimensions to be $\frac{2\pi^2}{1-\varepsilon}$\,.

\begin{equation}
\int dp\,(p^2)^{\lambda}=0 ~~~~{\rm for~any} ~\lambda
\label{equationA7}
\end{equation}

\begin{equation}
\int \frac{dp}{p^{2\alpha}(p^2+m^2)^{\beta}}=
\frac{i \pi^2~\Gamma(\alpha+\beta-2+\varepsilon)
\Gamma(2-\alpha-\varepsilon)}
{(m^2)^{\alpha+\beta-2+\varepsilon}(1-\varepsilon)\Gamma(\beta)}
\label{equationA8}
\end{equation}

\begin{equation}
\int \frac{dq}{q^{2\alpha}(p-q)^{2\beta}}=
\frac{i\pi^2 \Gamma(1-\varepsilon)\Gamma(\alpha+\beta-2+\varepsilon)
\Gamma(2-\alpha-\varepsilon) \Gamma(2-\beta-\varepsilon) }
{(p^2)^{\alpha+\beta-2+\varepsilon} \Gamma(\alpha)\Gamma(\beta)
\Gamma(4-\alpha-\beta-2\varepsilon)}
\label{equationA9}
\end{equation}

\begin{equation}
\int \frac{dq~q_{\mu}}{q^{2\alpha}(p-q)^{2\beta}}=
\frac{i\pi^2 ~p_{\mu}~\Gamma(1-\varepsilon)\Gamma(\alpha+\beta-2+\varepsilon)
\Gamma(3-\alpha-\varepsilon) \Gamma(2-\beta-\varepsilon) }
{(p^2)^{\alpha+\beta-2+\varepsilon} \Gamma(\alpha)\Gamma(\beta)
\Gamma(5-\alpha-\beta-2\varepsilon)}
\label{equationA10}
\end{equation}

\begin{multline}
\label{equationA11}
\int \frac{dq~q_{\mu}q_{\nu}}
{q^{2\alpha}(p-q)^{2\beta}} \ = \
\frac{i\pi^2 ~\Gamma(1-\varepsilon)\Gamma(\alpha+\beta-3+\varepsilon)
\Gamma(3-\alpha-\varepsilon) \Gamma(2-\beta-\varepsilon) }
{(p^2)^{\alpha+\beta-2+\varepsilon} \Gamma(\alpha)\Gamma(\beta)
\Gamma(6-\alpha-\beta-2\varepsilon)} \\
\times[(\alpha+\beta-3+\varepsilon)(3-\alpha-\varepsilon)p_{\mu}p_{\nu}
+\frac12\,(2-\beta-\varepsilon)g_{\mu \nu}p^2]
\end{multline}

\begin{multline}
\label{equationA12}
\int \frac{dq~q_{\mu}q_{\nu} q_{\lambda}}
{q^{2\alpha}(p-q)^{2\beta}} \ = \
\frac{i\pi^2 ~\Gamma(1-\varepsilon)\Gamma(\alpha+\beta-3+\varepsilon)
\Gamma(4-\alpha-\varepsilon) \Gamma(2-\beta-\varepsilon) }
{(p^2)^{\alpha+\beta-2+\varepsilon} \Gamma(\alpha)\Gamma(\beta)
\Gamma(7-\alpha-\beta-2\varepsilon)} \\
\times[(\alpha+\beta-3+\varepsilon)(4-\alpha-\varepsilon)
p_{\mu}p_{\nu}p_{\lambda}
+\frac12(2-\beta-\varepsilon)p^2 ( p_{\mu} g_{\nu \lambda}
+p_{\nu} g_{\mu \lambda} +p_{\lambda} g_{\mu \nu} )]
\end{multline}

\vspace{.5cm}

\noindent
\textbf{Two-loop integration formulas [17]}

\vspace{0cm}

$$
\frac{(p^2)^{\alpha+\beta+\gamma+\sigma+\rho-4+2\varepsilon}}{(i\pi^2)^2}
\int \!\frac{dt~dq}{t^{2\alpha} q^{2\beta} (p-t)^{2\gamma}
(p-q)^{2\sigma} (t-q)^{2\rho}} \ \equiv \ V(\alpha,\beta,\gamma,\sigma,\rho)\,.
$$
\begin{multline}
\label{equationA13}
V(\alpha,1,\gamma,1,1) \ = \ \frac{\Gamma^3(1-\varepsilon)\Gamma(-1+2\varepsilon)
 \Gamma(1-\alpha-\varepsilon)\Gamma(1-\gamma-\varepsilon)\Gamma(\alpha+\gamma-2+2\varepsilon)}
 {\Gamma(\alpha)\Gamma(\gamma)\Gamma(3-\alpha-\gamma-3\varepsilon)} \\
\times\left[\frac{\Gamma(3-\alpha-\gamma-3\varepsilon)}{\Gamma(2-\alpha-\gamma-\varepsilon)}
 - \frac{\Gamma(\alpha+\gamma-1+\varepsilon)}{\Gamma(\alpha+\gamma-2+3\varepsilon)}
 + \frac{\Gamma(\alpha)}{\Gamma(\alpha-1+2\varepsilon)}\right. \\
+ \left.\frac{\Gamma(\gamma)}{\Gamma(\gamma-1+2\varepsilon)}
 - \frac{\Gamma(2-\alpha-2\varepsilon)}{\Gamma(1-\alpha)}
 - \frac{\Gamma(2-\gamma-2\varepsilon)}{\Gamma(1-\gamma)}\right]
\end{multline}
\begin{multline}
\label{equationA14}
V(\alpha,\beta,1,1,\rho) \ = \ \frac{\Gamma^3(1-\varepsilon)\Gamma(2-\alpha-\varepsilon)
 \Gamma(2-\beta-\varepsilon)\Gamma(2-\rho-\varepsilon)}
 {\Gamma(2-2\varepsilon)\Gamma(\alpha)\Gamma(\beta)\Gamma(\rho)} \\
\times\sum^\infty_{m,n=0}\frac{(-)^m \Gamma(n+2-2\varepsilon)
 \Gamma(m+n+\alpha+\beta+\rho-2+2\varepsilon)}
{m! n! (n+1-\varepsilon)\Gamma(4-m-\alpha-\beta-\rho-3\varepsilon)\Gamma(m+n+2-\varepsilon)} \\
\times\left[\frac{1}{(n+\rho)(m+n+\alpha+\rho-1+\varepsilon)}
 + \frac{1}{(n+\rho)(m+n+\beta+\rho-1+\varepsilon)}\right. \\
\left. + \frac{1}{(m+n+\alpha)(m+n+\alpha+\rho-1+\varepsilon)}
 + \frac{1}{(m+n+\beta)(m+n+\beta+\rho-1+\varepsilon)}\right. \\
\left. + \frac{1}{(m+n+\alpha)(n+2-\rho-2\varepsilon)}
 + \frac{1}{(m+n+\beta)(n+2-\rho-2\varepsilon)}\right]
\end{multline}

\vspace{.5cm}

\noindent
\textbf{Individual two-loop integrals}

\vspace{.2cm}

Here we write down the relevant integrals $V(\alpha,\beta,\gamma,\sigma,\rho)$
with all the arguments being positive integers, retaining the
$\frac{1}{\varepsilon^2},\frac{1}{\varepsilon}$ and $O(1)$ terms.
\begin{align*}
V(1,1,1,1,1) \ &= \ 6\zeta(3) \\
V(2,1,1,1,1) \ &= \ \frac{1}{2\varepsilon^2} - \frac{1}{2\varepsilon} + \frac12 \\
V(1,1,1,1,2) \ &= \ \frac{1}{\varepsilon^2} + \frac{1}{\varepsilon} - 3 \\
V(2,2,1,1,1) \ &= \ \frac{1}{\varepsilon} - \frac52 \\
V(2,1,2,1,1) \ &= \ \frac{1}{\varepsilon^2} - \frac{1}{\varepsilon} - 1 \\
V(2,1,1,2,1) \ &= \ \frac{2}{\varepsilon^2} + \frac{3}{\varepsilon} - 1 \\
V(3,1,1,1,1) \ &= \ \frac{1}{4\varepsilon^2} + \frac{5}{8\varepsilon} + \frac{11}{16}
\end{align*}

\vspace{.5cm}

\noindent
\textbf{Pole parts of the essentially three-loop integrals of the form}
$$
\frac{(k^2)^{3\varepsilon-1}}{(i\pi^2)^3}\,\int\frac{dp~dq~dt\ \,Y(p,q,t,k)}
{p^2 q^2 t^2 (k-p)^2 (k-q)^2 (k-t)^2 (p-q)^2 (p-t)^2 (q-t)^2}\,.
$$
\begin{align*}
Y = \ (p-t)^8 \ \ &\Longrightarrow \ \ - \frac{2}{3\varepsilon^3}
 - \frac{61}{18\varepsilon^2} - \frac{877}{108\varepsilon}
 + \frac{4}{\varepsilon}\,\zeta(3) \\
(p-t)^6 k^2 \ \ &\Longrightarrow \ \ \frac{1}{\varepsilon^3}
 + \frac{41}{6\varepsilon^2} + \frac{31}{\varepsilon}
 - \frac{6}{\varepsilon}\,\zeta(3) \\
(p-t)^4 k^4 \ \ &\Longrightarrow \ \ \frac{12}{\varepsilon}\,\zeta(3) \\
(k-q)^8 \ \ &\Longrightarrow \ \ \frac{2}{3\varepsilon^2}
 + \frac{49}{6\varepsilon} + \frac{4}{\varepsilon}\,\zeta(3) \\
(k-q)^6 k^2 \ \ &\Longrightarrow \ \ \frac{1}{3\varepsilon^2}
 + \frac{4}{\varepsilon} + \frac{4}{\varepsilon}\,\zeta(3) \\
(k-q)^4 k^4 \ \ &\Longrightarrow \ \ \frac{4}{\varepsilon}\,\zeta(3) \\
(k-q)^2 k^6 \ \ &\Longrightarrow \ \ -\frac{2}{\varepsilon}\,\zeta(3) \\
(k-q)^4 (p-t)^4 \ \ &\Longrightarrow \ \ \frac{1}{2\varepsilon^2}
 + \frac{17}{3\varepsilon} \\
(k-q)^6 (p-t)^2 \ \ &\Longrightarrow \ \ \frac{5}{12\varepsilon^3}
 + \frac{73}{24\varepsilon^2} + \frac{661}{48\varepsilon} \\
(k-q)^2 (p-t)^6 \ \ &\Longrightarrow \ \ -\frac{1}{4\varepsilon^3}
 - \frac{65}{24\varepsilon^2} - \frac{865}{48\varepsilon} \\
(k-q)^4 (p-t)^2 k^2 \ \ &\Longrightarrow \ \ \frac{1}{3\varepsilon^3}
 + \frac{7}{3\varepsilon^2} + \frac{31}{3\varepsilon} \\
(k-q)^2 (p-t)^4 k^2 \ \ &\Longrightarrow \ \ \frac{1}{3\varepsilon^3}
 + \frac{3}{\varepsilon^2} + \frac{53}{3\varepsilon} \\
(k-q)^4 p^4 \ \ &\Longrightarrow \ \ \frac{1}{6\varepsilon^3}
 + \frac{17}{12\varepsilon^2} + \frac{199}{24\varepsilon} \\
(k-q)^6 p^2 \ \ &\Longrightarrow \ \ \frac{1}{8\varepsilon^3}
 + \frac{49}{48\varepsilon^2} + \frac{531}{96\varepsilon} \\
(k-q)^4 p^2 k^2 \ \ &\Longrightarrow \ \ \frac{1}{6\varepsilon^3}
 + \frac{3}{2\varepsilon^2} + \frac{55}{6\varepsilon}
\end{align*}

\vspace{.5cm}

\begin{center}
REFERENCES
\end{center}

\begin{enumerate}
\item
\textit{Buras A.\,J.} \,
Asymptotic Freedom in Deep Inelastic Processes in the Leading Order
and Beyond // Rev.\,Mod.\,Phys. 1980. V.\,52. P.\,199-276.
\item
\textit{Le Guillou J.\,C., Zinn-Justin J.} \,
Critical Exponents for the $n$-Vector Model in
Three Dimensions from Field Theory //
Phys.\,Rev.\,Lett. 1977. V.\,39. P.\, 95-98; \\
\textit{Kazakov D.\,I., Tarasov O.\,V., Vladimirov A.\,A.}
Calculation Of Critical Exponents By Quantum Field Theory Methods //
Sov.\,Phys.\,JETP. 1979. V.\,50. P.\,521-526.
\item
\textit{Brink L., Schwarz J.\,H., Scherk J.} \,
Supersymmetric Yang-Mills Theories //
Nucl. Phys.\,B. 1977. V.\,121. P.\,77-92.
\item
\textit{Gliozzi F., Scherk J., Olive D.} \,
Supersymmetry, Supergravity Theories and the Dual Spinor Model //
Nucl.\,Phys.\,B. 1977. V.\,122. P.\,253-290.
\item
\textit{Jones D.\,R.\,T.} \,
Charge Renormalization in a Supersymmetric Yang-Mills Theory //
Phys.\,Lett.\,B. 1977. V.\,72. P.\,199; \\
\textit{Poggio E.\,C., Pendleton H.\,N.} \,
Vanishing of Charge Renormalization and Anomalies in a Supersymmetric
Gauge Theory //
Phys.\,Lett.\,B. 1977. V.\,72. P.\,200-202.
\item
\textit{Chetyrkin K.\,G., Kataev A.\,L., Tkachov F.\,V.} \,
Higher Order Corrections to Sigma-t (e+ e- ---$>$ Hadrons) in
Quantum Chromodynamics //
Phys.\,Lett.\,B. 1979. V.\,85. P.\,277-279.
\item
\textit{Dine M., Sapirstein J.} \,
Higher Order QCD Corrections in e+ e- Annihilation //
Phys.\,Rev.\,Lett. 1979. V.\,43. P.\,668-671.
\item
\textit{Celmaster W., Gonsalves R.\,J.} \,
An Analytic Calculation of Higher Order Quantum Chromodynamic Corrections
in e+ e- Annihilation //
Phys.\,Rev.\,Lett. 1980. V.\,44. P.\,560-564.
\item
\textit{Caswell W.\,E.} \,
Asymptotic Behavior of Nonabelian Gauge Theories to Two Loop Order //
Phys.\,Rev.\,Lett. 1974. V.\,33. P.\,244-246; \\
\textit{Jones D.\,R.\,T.} \,
Two Loop Diagrams in Yang-Mills Theory //
Nucl.\,Phys.\,B. 1974. V.\,75. P.\,531-538.
\item
\textit{Slavnov A.\,A.} \,
Ward Identities in Gauge Theories //
Theor.\,Math.\,Phys. 1972. V.\,10. P.\,99-107; \\
\textit{Taylor J.\,C.} \,
Ward Identities and Charge Renormalization of the Yang-Mills Field //
Nucl.\,Phys.\,B. 1971. V.\,33. P.\,436-444.
\item
\textit{'t\,Hooft G.} \,
Dimensional regularization and the renormalization group //
Nucl. Phys.\,B. 1973. V.\,61. P.\,455-468; \\
\textit{Collins J.\,C., Macfarlane A.\,J.} \,
New methods for the renormalization group //
Phys.\,Rev.\,D. 1974. V.\,10. P.\,1201-1212.
\item
\textit{Caswell W.\,E., Wilczek F.} \,
On the Gauge Dependence of Renormalization Group Parameters //
Phys.\,Lett.\,B. 1974. V.\,49. P.\,291-292; \\
\textit{Kallosh R.\,E., Tyutin I.\,V.} \,
The Gauge Invariance of the Renormalization Group Equations //
Sov.\,J.\,Nucl.\,Phys. 1975. V.\,20. P.\,653-656.
\item
\textit{Egorian E., Tarasov O.\,V.} \,
Two Loop Renormalization Of The QCD In An Arbitrary Gauge //
Theor.\,Math.\,Phys. 1979. V.\,41. P.\,863-867.
\item
\textit{Vladimirov A.\,A.} \,
Methods of Multiloop Calculations and the Renormalization Group
Analysis of phi**4 Theory //
Theor.\,Math.\,Phys. 1979. V.\,36. P.\,732-737.
\item
\textit{Vladimirov A.\,A.} \,
Method For Computing Renormalization Group Functions In Dimensional
Renormalization Scheme //
Theor.\,Math.\,Phys. 1980. V.\,43. P.\,417-422.
\item
\textit{Speer E.\,R.} \,
Renormalization and Ward identities using complex space-time dimension //
J.\,Math.\,Phys. 1974. V.\,15. P.\,1-6; \\
\textit{Collins J.\,C.} \,
Structure of Counterterms in Dimensional Regularization //
Nucl. Phys.\,B. 1974. V.\,80. P.\,341-348; \\
\textit{Breitenlohner P., Maison D.} \,
Dimensional Renormalization and the Action Principle //
Commun.\,Math.\,Phys. 1977. V.\,52. P.\,55-75.
\item
\textit{Chetyrkin K.\,G., Tkachov F.\,V.} \,
A New Approach To Evaluation Of Multiloop Feynman Integrals.
Preprint INR $\Pi$-0118. Moscow, 1979. 12 p.
\item
\textit{Cvitanovic P.} \,
Group Theory for Feynman Diagrams in Nonabelian Gauge Theories:
Exceptional Groups //
Phys.\,Rev.\,D. 1976. V.\,14. P.\,1536-1553.
\item
\textit{Strubbe H.} \,
Manual for Schoonschip: A CDC 6000/7000 program for symbolic evaluation
of algebraic expressions //
Comput.\,Phys.\,Commun. 1974. V.\,8. P.\,1-30.
\item
\textit{Vladimirov A.\,A., Shirkov D.\,V.} \,
The Renormalization Group And Ultraviolet Asymptotics //
Sov.\,Phys.\,Usp. 1979. V.\,22. P.\,860-878.
\item
\textit{Curtright T., Ghandour G.} \,
Stability and Supersymmetry: General Formalism and Explicit
Two Loop Applications //
Annals Phys. 1977. V.\,106. P.\,209-278; \\
\textit{Townsend P.\,K., van Nieuwenhuizen P.} \,
Dimensional Regularization And Supersymmetry At The Two Loop Level //
Phys.\,Rev.\,D. 1979. V.\,20. P.\,1832-1838; \\
\textit{Sezgin E.} \,
Dimensional Regularization And The Massive Wess-Zumino Model //
Nucl.\,Phys.\,B. 1980. V.\,162. P.\,1-11; \\
\textit{Siegel W.} \,
Supersymmetric Dimensional Regularization via Dimensional Reduction //
Phys.\,Lett.\,B. 1979. V.\,84. P.\,193-196.
\item
\textit{Abbott L.\,F., Grisaru M.\,T., Schnitzer H.\,J.} \,
Supercurrent Anomaly in a Supersymmetric Gauge Theory //
Phys.\,Rev.\,D. 1977. V.\,16. P.\,2995-3001; \\
\textit{Curtright T.} \,
Conformal Spinor Current Anomalies //
Phys.\,Lett.\,B. 1977. V.\,71. P.\,185-188.
\end{enumerate}

\end{document}